\documentclass[fleqn,usenatbib]{mnras}

\usepackage{newtxtext,newtxmath}
\usepackage[T1]{fontenc}

\DeclareRobustCommand{\VAN}[3]{#2}
\let\VANthebibliography\thebibliography
\def\thebibliography{\DeclareRobustCommand{\VAN}[3]{##3}\VANthebibliography}

\usepackage{graphicx}	% Including figure files
\usepackage{amsmath}	% Advanced maths commands
\usepackage{xcolor}
\usepackage{longtable}
\usepackage{multirow}

\newcommand{\WD}{SDSS\,J1637+3631}
\newcommand{\WDlong}{SDSS\,J163712.21+363155.9}
\newcommand{\Dsix}{D$^6$}
\newcommand{\Teff}{\mbox{$T_\mathrm{eff}$}}
\newcommand{\Msun}{\mbox{$\mathrm{M}_\odot$}}
\newcommand{\Rsun}{\mbox{$\mathrm{R}_\odot$}}
\newcommand{\Lsun}{\mbox{$\mathrm{L}_\odot$}}

\newcommand{\Ion}[2]{#1\,\textsc{#2}}
\newcommand{\kms}{\mbox{km\,s$^{-1}$}}
\newcommand{\ebv}{\mbox{$E(B-V)$}}
\newcommand{\lp}{\mbox{LP\,40$-$365}}
\newcommand{\dd}{\mathrm{d}}
\newcommand{\logXY}[2]{\mbox{$\log(\mathrm{#1}/\mathrm{#2})$}}
\newcommand{\logXC}[1]{\mbox{$\log(\mathrm{#1}/\mathrm{C})$}}

\title[Analysis of a warm \Dsix\ survivor]{Spectroscopic and kinematic analyses of a warm survivor of a \Dsix\ supernova}

\author[M. A. Hollands et al.]{
M. A. Hollands,$^{1}$\thanks{E-mail: mark.hollands@warwick.ac.uk}
K. J. Shen,$^{2}$
R. Raddi$^{3}$,
B. T. G\"ansicke,$^{1}$
E. B. Bauer,$^{4}$
and
A. Rebassa-Mansergas$^{3,5}$
\\
$^{1}$ Department of Physics, University of Warwick, Coventry, CV4 7AL, UK \\
$^{2}$ Department of Astronomy and Theoretical Astrophysics Center, University of California, Berkeley, CA 94720, USA \\
$^{3}$ Departament de F\'isica, Universitat Polit\`ecnica de Catalunya, c/Esteve Terrades 5, 08860 Castelldefels, Spain \\
$^{4}$ Center for Astrophysics | Harvard \& Smithsonian, 60 Garden St., Cambridge, MA 02138, USA\\
$^{5}$ Institut d'Estudis Espacials de Catalunya, Esteve Terradas, 1, Edifici RDIT, Campus PMT-UPC, 08860 Castelldefels, Barcelona, Spain
}

\date{Accepted 2025 June 09. Received 2025 June 09; in original form 2025 April 11}

\pubyear{2025}

\begin{document}
\label{firstpage}
\pagerange{\pageref{firstpage}--\pageref{lastpage}}
\maketitle

\begin{abstract}
\WDlong\ is a candidate hyper-runaway star, first identified from its unusual
spectrum in the Sloan Digital Sky Survey, which exhibits oxygen, magnesium, and
silicon lines redshifted by several 100\,\kms, leading to the suggestion it was
ejected from a thermonuclear supernova. We have acquired GTC OSIRIS
spectroscopy of \WD\ establishing a warm ($\Teff=15\,680\pm250$\,K)
carbon+oxygen dominated atmosphere, that is also abundant in the intermediate
mass elements silicon, sulphur, and calcium. We interpret \WD\ as the donor to
an accreting white dwarf that exploded in a dynamically-driven
double-degenerate double-detonation (\Dsix) type Ia supernova, where the
current composition is consistent with a CO white dwarf core, enriched with
intermediate mass elements from deposited supernova ejecta. While \WD\ has a
low-precision \emph{Gaia} parallax, our spectroscopic surface gravity ($\log
g=6.3\pm0.3$\,dex) helps constrain its tangential velocity to
$1950_{-530}^{+810}$\,\kms, providing additional support to the \Dsix\
mechanism. Under the assumption that \WD\ is a \Dsix\ survivor, we construct a
kinematic model combining all astrometric, spectroscopic, and photometric
information, but also including the structure and gravitational potential of
the Milky Way. Our model localises the ejection site to the inner few kpc of
the Galactic disc (though excluding the Galactic centre), with an ejection
speed of $1870_{-300}^{+360}$\,\kms, and a $4.5_{-0.5}^{+0.4}$\,Myr time of
flight.
\end{abstract}


%%%%%%%%%%%%%%%%%%%%%%%%%%%%%%%%%%%%%%%%%%%%%%%%%%

%%%%%%%%%%%%%%%%% BODY OF PAPER %%%%%%%%%%%%%%%%%%

\section{Introduction}

Type Ia supernovae result from the thermonuclear detonation of a white dwarf,
the stellar remnant left behind by the vast majority of main sequence stars.
For a white dwarf to undergo such a destructive process requires a high enough
density to trigger carbon fusion within the interior \citep{khokhlov91-1}.
These conditions can be achieved when the white dwarf is part of a multiple
star system, either by accretion from or by a merger with a companion (where
the companion may also be a white dwarf). As astrophysical transients, type Ia
supernovae can easily outshine their host galaxies, permitting their use as
standardisable candles for measuring cosmic distances, which led to the
discovery that the expansion of the universe is accelerating
\citep{riessetal98-1,perlmutteretal99-1}. Despite their importance in
cosmology, several open questions remain regarding type Ia supernovae,
primarily the specific channels that lead to their occurrence
\citep{wang+han12-1,ruiter+seitenzahl25-1}, the relative rates at which these
channels occur, and how the yields influence the chemical evolution of the
universe \citep{thielemannetal86-1}.

Over the last decade it has become clear that in some cases, white dwarfs or
their companions can survive type Ia supernovae, providing new avenues to study
them. Evidence for such supernova survivors has come from extragalactic
observations of peculiar supernovae \citep{mccullyetal14-1}, theoretical
predictions from supernova modelling, and finally the observation of chemically
and/or kinematically peculiar stars within the Milky Way galaxy. Such peculiar
stars that have been found within the Milky Way are reviewed below.

The first Galactic star proposed to be be the survivor of a supernova is
US\,708, which was found by \citet{geieretal15-1} to be a helium-rich hot
subdwarf moving faster than 1000\,\kms, making it the fastest known unbound
stellar object in the Galaxy at that time. Hypervelocity stars, i.e. stars
unbound to the Milky Way, were previously understood to be ejections from the
Galactic centre \citep{hills88-1,koposovetal20-1}. Although US\,708 lacked a
parallax at that time, the other proper-motion and radial velocity were
sufficient to show that a Galactocentric origin was highly improbable. This,
along with the discovery that US\,708 is a rapid rotator led
\citet{geieretal15-1} to conclude that this subdwarf was the donor to a white
dwarf which exploded as a type Ia supernova, where the donor survived, and was
ejected from the disrupted binary system.

Observations of extragalactic transients identified a class of peculiar
underluminous type Ia \citep{lietal03-1,jhaetal06-1}, termed type Iax
\citep{foleyetal13-1}. These transients were proposed to originate from the
deflagration of a white dwarf accreting from a helium-rich companion (e.g.
similar to US\,708). Simulations of such deflagrations were found to leave
behind a bound-remnant \citep{kromeretal13-1,finketal14-1,jonesetal19-1},
leading to the possibility that partially burned stars could exist in the Milky
Way and may be identified by their unusual compositions and kinematics. The
first object discovered fitting this description was \lp\ (GD\,492) by
\citet{vennesetal17-1}. \lp\ is found to have a neon dominated atmosphere and
has a spectrum populated with metal lines from a variety of elements, as well
as a radial velocity of $\simeq 500$\,\kms. The resemblance to an exposed ONe
white dwarf core, potentially enhanced in supernova burning products with a
super-solar Mn/Fe ratio \citep{raddietal18-1}, indicated a near-Chandrasekhar
mass deflagration \citep{seitenzahletal13-2}. With the release of \emph{Gaia}
data release 2 (DR2), \lp\ could be placed on the Hertzsprung-Russell diagram
(HRD), where it was found to reside between the white dwarf and main sequences
\citep{raddietal18-2}, implying a low mass and inflated radius compared to
normal white dwarfs, but more compact than main sequence stars of a similar
temperature. Additional objects in the same spectral class as \lp\ have since
been identified with similar compositions and high-speed kinematics
\citep{raddietal19-1,el-badryetal23-1}. 

One process predicted to lead directly to standard type Ia supernovae is the
double-detonation mechanism \citep{taam80-1,livne90-1}. In a double-detonation
supernova, an accreting white dwarf primary (which may have a mass well below
the Chandrasekhar-limit; \citealt{finketal07-1,finketal10-1}) undergoes an
initial detonation in its helium envelope \citep{shenetal09-1}, where the
helium burning shock front travels around the stellar surface and is focussed
onto the opposite side, triggering a second detonation of the carbon in the
white dwarf core \citep{shen+bildsten14-1}. In double-degenerate systems, where
the Roche-lobe filling donor star is also a white dwarf (which could harbour
either a He core or a CO core with a He-rich envelope), double detonations are
expected to occur
\citep{guillochonetal10-1,danetal11-1,raskinetal12-1,pakmoretal13-1}, and can
potentially lead to another kind of supernova survivor: Although the primary
white dwarf is destroyed, the donor may survive interaction with the supernova
shock, being flung out of the system \citep{shenetal18-1} with approximately
its orbital velocity. This type of supernova has been termed the
dynamically-driven double-degenerate double-detonation (\Dsix) mechanism, and
can potentially occur in all CO-core white dwarfs with masses below 1\,\Msun\
\citep{shenetal24-1}. While double detonations offer the potential to explain a
large fraction of all type Ia supernovae \citep{ruiter+seitenzahl25-1},
searches within/around supernova remnants for surviving donors have proved
unsuccessful \citep{shenetal18-1,shieldsetal23-1}. Another challenge for the
\Dsix\ mechanism is matching the estimated Galactic supernova rate to the
number of known survivors \citep{el-badryetal23-1}. However, recent work has
shown that interaction between the ejecta and donor can trigger a further
double detonation, destroying both stars,
\citep{tanikawaetal19-1,pakmoretal22-1,boosetal24-1}, and with \citet{shen25-1}
indicating this occurs in the majority of cases, where only a few percent of
systems leave behind surviving donors.

A defining observational signature for the survivors of \Dsix\ supernovae,
simply referred to as \Dsix\ stars, is their huge ejection velocities which are
expected between 1000 to 3000\,\kms\ \citep{shenetal18-1}. Having interacted
with the high-velocity supernova ejecta, the donor is expected to have
undergone mass loss due to stripping of its envelope, which may further be
chemically enhanced by ejecta material accreted during the explosion. \Dsix\
stars are therefore also expected to have unusual spectra that may be dominated
with carbon and oxygen from the exposed white dwarf core, but enhanced with
traces of heavier metals.

With the release of \emph{Gaia} DR2, \citet{shenetal18-1}, searched for stars
with extreme tangential velocities ($v_\perp$) exceeding 1000\,\kms, following
up with low-resolution spectroscopy. Of their seven candidates, four turned out
to be ordinary stars with probable underestimated parallaxes. However, the
remaining three (named D6-1, D6-2, and D6-3) had unusual (but mutually similar)
spectra dominated by atomic lines from oxygen and carbon C$_2$ molecular bands,
as well as metal lines from magnesium and calcium. All three had estimated
$\Teff \approx 8000$\,K, and their sharp \Ion{Ca}{ii} H+K lines established
D6-1 to have a radial velocity of 1200\,\kms -- though curiously D6-2 and D6-3
had radial velocities consistent with zero. The past trajectory of D6-2 was
also found to coincide with the supernova remnant G70.0-21.5. Evidently these
stars match the expected properties of \Dsix\ survivors, though to date are yet
to receive dedicated spectral analyses of their atmospheres (Hollands et al.,
in prep). Similar searches for high $v_\perp$ objects with precise measurements
in \emph{Gaia} DR3 did not reveal any new \Dsix\ candidates
\citep{igoshevetal23-1}, and constructing clean searches for hyper-runaway
stars has been shown to be a complex task \citep{scholz24-1}.

With the lowest hanging fruit already discovered by \citet{shenetal18-1},
\citet{el-badryetal23-1} identified four more \Dsix\ stars by searching for
blue objects with low parallax significance, but still with sizeable
proper-motions. Unlike the cool stars found by \citet{shenetal18-1}, these new
discoveries were much hotter, with most having $\Teff > 60\,000$\,K. The
atmospheres of these hottest stars were first analysed by
\citet{werneretal24-1} identifying carbon and oxygen rich atmospheres for two
of them, with a third having a more typical DAO white dwarf spectrum. One of
these objects, J0927$-$6335, has been followed up with ultraviolet observations
\citep{werneretal24-2}, establishing an atmosphere not only rich in carbon and
oxygen, but also enhanced in silicon, iron and nickel. However, the authors
acknowledge that the abundances of these hot \Dsix\ stars may be affected by
radiative levitation, and that the effect of gravitational settling is not
known for these stars.

\subsection{\texorpdfstring{\WD}{SDSS J1637+3631}}

Following the second data release of \emph{Gaia}, \citet{raddietal19-1}
presented the discovery of additional stars in the same class as \lp. Two
objects were identified from their position in the HRD, large tangential
velocities, and subsequent follow-up spectroscopy. Having found these stars to
have distinctive spectra dominated by strong magnesium and oxygen lines, the
authors searched the Sloan Digital Sky Survey (SDSS) spectroscopic database for
additional \lp-like candidates that may have been observed serendipitously.
Using a grid of \lp-like templates at different \Teff\ and radial velocity, the
authors identified three objects with peculiar spectra.

The first of these was a known white dwarf \citep{kepleretal16-1},
SDSS\,J1240$+$6710, which has a unique oxygen dominated atmosphere, and has
been suggested to have formed from a thermonuclear burning event distinct from
the mechanisms producing the \Dsix\ and \lp\ remnants
\citep{gaensickeetal20-1}.

The second object, SDSS\,J0905$+$2510, was found to closely match one of the
\lp\ templates, adding a fourth member to the class of \lp\ stars known at that
time. SDSS\,J0905$+$2510 could not have been identified using the astrometric
approach, as its \emph{Gaia} DR2 record did not have an available parallax or
proper-motion.\footnotemark

\footnotetext{J0905$+$2510 now has full 5-parameter astrometry in \emph{Gaia}
DR3, though its parallax is measured to less than $1\sigma$ precision
($0.2879\pm0.4987$\,mas).}

The final peculiar object found by \citep{raddietal19-1} using their template
fitting approach is the star \WDlong\ (\WD\ hereafter), and is the primary
subject of this work. \WD\ appeared to be hotter than the other objects
analysed by \citet{raddietal19-1}, best-matching the hottest template in their
grid ($20\,000$\,K). The star was found to match several velocity-shifted
($\approx 300$\,\kms) metal lines in the template, arising from \Ion{Mg}{ii},
\Ion{O}{i}, and \Ion{Si}{ii}. Even so, other lines were found to be present in
the data that were either absent or much weaker in the template, such as the
\Ion{Ca}{ii} H+K lines, and further \Ion{O}{i} lines. \WD, was therefore
considered unlikely to originate from the same mechanism as the other stars in
the \lp\ class, and may instead belong to one of the other exotic classes of
remnants related to white dwarf supernovae, such as the aforementioned
oxygen-rich white dwarf SDSS\,J1240$+$6710, or surviving donors of the \Dsix\
supernovae. However, its origin could not be revealed from its \emph{Gaia}
astrometry, as its DR2 parallax of $0.39\pm0.60$\,mas (now revised to
$0.33\pm0.48$\,mas in DR3: see Table~\ref{tab:astromphot} for all DR3
astrometry and available optical photometry) was insufficiently precise to
establish either its location in the \emph{Gaia} HRD or determine the magnitude
of its tangential velocity, $v_\perp$ (though the 67\,mas\,yr$^{-1}$ hinted at
a fast moving star).

\begin{table}
    \centering
    \begin{tabular}{l|c}
        \hline
        Parameter & value \\
\hline
Gaia DR3 source ID & 1327920737357113088 \\
RA (J2016) & 16:37:12.214 \\
Dec (J2016) & +36:31:55.91 \\
$\varpi$ [mas]     &  $0.3319\pm0.4776$ \\
$\mu_\mathrm{RA}$  [mas\,yr$^{-1}$] & $-19.558\pm0.587$ \\
$\mu_\mathrm{Dec}$ [mas\,yr$^{-1}$] & $+64.159\pm0.644$ \\
\hline
\emph{Gaia} $G$ & $20.2549\pm0.0063$ \\
\emph{Gaia} $G_\mathrm{BP}$ & $20.0409\pm0.0698$ \\
\emph{Gaia} $G_\mathrm{RP}$ & $20.1670\pm0.1319$ \\
SDSS $u$ & $20.008\pm0.053$\\
SDSS $g$ & $20.092\pm0.021$\\
SDSS $r$ & $20.399\pm0.039$\\
SDSS $i$ & $20.623\pm0.064$\\
SDSS $z$ & $20.861\pm0.282$\\
PanSTARRS $g$ & $20.1335\pm0.0108$ \\
PanSTARRS $r$ & $20.4299\pm0.0127$ \\
PanSTARRS $i$ & $20.6833\pm0.0287$ \\
PanSTARRS $z$ & $20.8248\pm0.0389$ \\
\hline
    \end{tabular}
    \caption{Astrometry and photometry for \WD.}
    \label{tab:astromphot}
\end{table}

In this work, we provide new spectroscopic observations and analysis of \WD,
establishing that this elusive object is the ejected donor from a \Dsix\
supernova. In section~\ref{sec:obs}, we present our spectroscopic observations
of \WD. In section~\ref{sec:spec} we perform a detailed spectral analysis of
our observations, establishing a C+O dominated composition enhanced with
heavier elements such as Si and Ca. In section~\ref{sec:confD6}, we combine
several pieces of information to conclude that \WD\ is the surviving donor from
a D$^6$ explosion. In section~\ref{sec:kin}, we take these arguments further to
determine the kinematic properties of \WD, its likely birth site, and
time-of-flight. Finally, we present our conclusions in section~\ref{sec:conc}.

\section{Spectroscopic observations}
\label{sec:obs}

We obtained follow up spectroscopy of \WD\ using the OSIRIS (Optical System for
Imaging and low-Intermediate-Resolution Integrated Spectroscopy) instrument
mounted on the Gran Telescopio CANARIAS (GTC) at the Roque de los Muchachos
observatory on La Palma. These observations were performed in service mode on
the night starting 2019 June 08 under photometric observing conditions and
seeing of 0.7\,arcsec. To obtain broad spectral coverage across the optical at
intermediate resolution, we used three instrumental setups with the R2500U,
R2500V, and R2500R VPH gratings, all with a 0.6\,arcsec slit width. This gave
continuous spectral coverage over the range 3440--7685\,\AA, with a resolving
power of $R\approx2500$ in all three gratings. For both the R2500U and R2500V
gratings, 1 hour of integration time was acquired over three sub-exposures each
(i.e. 1200\,s per exposure). For the R2500R grating, we opted for six 1200\,s
exposures totalling 2 hours of integration time. The additional exposure time
for R2500R was chosen because \WD\ is intrinsically blue and the detection of
neon (a key element in the atmospheres of SDSS\,J1240+6710 and the \lp\ stars)
from weak \ion{Ne}{i} lines (6144/6404/6508\,\AA\ being the strongest) would
depend on having a higher spectral signal-to-noise ratio (at least 30) than
would be necessary at bluer wavelengths.

Standard calibration data were acquired with flat field frames and arc lamp
exposures (for all three gratings) taken at the end of the night. The provided
bias frames were taken at the start of the following night (2019-06-09). The DA
white dwarf GD\,153 was observed as a spectral flux standard shortly before our
science observations were taken and at a similar airmass to the target
($\approx 1.05$).

\begin{figure*}
    \centering
    \includegraphics[width=\textwidth]{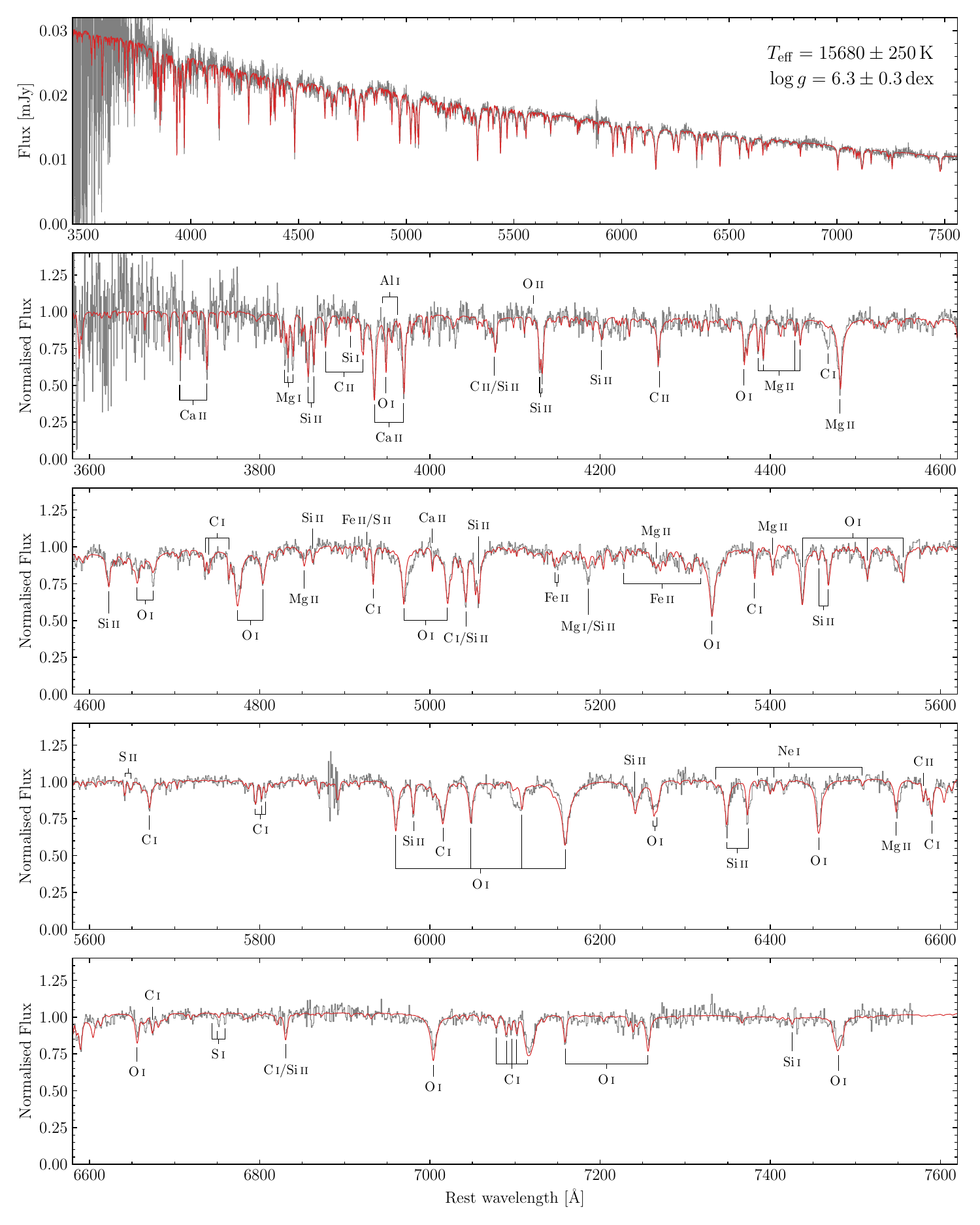}
    \caption{GTC OSIRIS spectrum of \WD\ (grey) with best fitting model (red).
    The top panel shows the entire GTC dataset in physical flux units.
    Subsequent panels show portions of the spectral range with fluxes
    normalised to one, and with labelled spectral features. All panels are in
    vacuum wavelengths with the data shifted to the rest frame.}
    \label{fig:GTCspec}
\end{figure*}

We performed reduction of the data using the \textsc{starlink} package of
software \citep{currieetal14-1,belletal24-1}, which included bias subtraction,
flat fielding, and optimal extraction \citep{horne86-1,marsh89-1} of the
spectra with each grating. Wavelength and flux calibrations (including telluric
removal) were performed using \textsc{molly}.\footnotemark\ Sub-exposures of
each band were combined by weighted averaging, with the three bands merged into
a single spectrum, averaging similarly over the overlapping regions. The final
coadded spectrum is shown in Figure~\ref{fig:GTCspec} (top panel).

\footnotetext{\textsc{molly} can be found at
\url{https://cygnus.astro.warwick.ac.uk/phsaap/software/}}

\section{Spectroscopic analysis}
\label{sec:spec}

Our new GTC spectrum allowed us to identify many more elements in the
atmosphere of \WD\ than was possible with the SDSS spectrum, owing to the much
higher signal-to-noise ratio. In addition to the previously identified lines of
\Ion{O}{i}, \Ion{Mg}{ii}, \Ion{Si}{ii}, and \Ion{Ca}{ii} \citep{raddietal19-1},
we also identified lines from \Ion{C}{i--ii}, \Ion{O}{ii}, \Ion{Ne}{i},
\Ion{Al}{i}, \Ion{Si}{i}, \Ion{S}{i--ii}, and \Ion{Fe}{ii} (see labelled panels
in Figure~\ref{fig:GTCspec}). No lines of hydrogen or helium were detected.
Arguably the most important of these identifications is carbon, which despite
being undetectable in the SDSS spectrum, shows many strong lines and blends
(such as the deep blend near a rest wavelength of 7115\,\AA) indicating this
element constitutes a large proportion of the atmosphere. Like in the
\lp-stars, neon is detected, but despite being much hotter (leading to higher
excitation of optical \Ion{Ne}{i} transitions), the lines are not as strong as
in the Ne-dominated atmospheres of the \lp-stars, indicating that this element
is not as abundant in \WD. In addition to these newly detected elements, the
spectrum allowed us to measure a radial velocity of $384\pm6$\,\kms, confirming
the fast kinematics suggested by the template fitting of the SDSS spectrum.

To model our GTC spectrum of \WD, we used the latest version of the Koester
white dwarf model atmosphere code \citep{koester10-1}. Compared to normal
metal-rich white dwarfs, which may only require the inclusion of hundreds to
thousands of metal lines, the apparently pure metal atmosphere of \WD\ required
tens of thousands of lines from all detected elements. Owing to the warm \Teff\
of \WD (a preliminary fit indicated a value of $\approx 16\,000$\,K), more than
half of the integrated flux is emitted at ultraviolet wavelengths, a region
which is densely populated with metal transitions. While the majority of these
lines are narrow (full width half maxima of a few 0.01\,\AA) and therefore
contribute negligible opacity on an individual basis, there are so many
thousands of such lines (of which a large fraction are saturated in the line
cores, though still almost vanish when convolved to the instrumental
resolution), that excluding these weak lines en masse would still cause a
drastic reduction in the total atmospheric opacity. Therefore all lines used in
this analysis were included in both the atmospheric structure calculation and
spectral synthesis. As a result, our calculations had to be evaluated on a
dense wavelength grid in order to properly sample the lines at all stages of
the model calculation. For the continuous opacity, we included bound-free
photoionisation cross-sections \citep{cuntoetal93-1} of C, O, Ne, Mg, Al, Si,
S, Ca, and Fe, and with all but Ne including the first two ionisation stages.
Additionally, the Koester models automatically include bound-free and free-free
opacities for the negative ions C$^-$ and O$^-$, and the free-free opacity for
Ne$^-$, when those elements are present in the calculation. However at the
temperature of \WD\, these negative ions were found to have a negligible
contribution compared to the aforementioned bound-free cross-sections. While we
did test our models with molecular abundances included, at this temperature the
most abundant molecular species: CO$^+$, O$_2^+$ and CO were present at number
fractions below $10^{-7}$. As such, including them had a negligible effect on
the atomic number abundances nor on the emergent flux from molecular opacities.
Therefore molecules were excluded from our fitting to reduce the computation
time. All fluxes in the atmospheric structure calculation were converged to 0.1
per\,cent accuracy.

Because of the large number of free parameters, and because the models were
expensive to compute compared to typical white dwarf atmospheres, it was
infeasible to construct multidimensional model grids for our fitting procedure.
We therefore performed a least squares fit to the spectrum with $\Teff$, $\log
g$, and the chemical abundances (relative to carbon) as free parameters,
recalculating the model at each step in the procedure. At each step in the fit,
the model was velocity shifted to the previous measurement of 384\,\kms, and
convolved to the instrumental resolving power of $2500$. To scale the model to
the observed fluxes we took the ratio of the data and model and fit a 5th order
polynomial across the entire wavelength range, then multiplying the model by
this wavelength dependent polynomial. This procedure accounts for the unknown
solid angle, interstellar extinction, and uncertainty in flux calibration. 

While we quickly established that oxygen is the most abundant element by
number, we found that the bound-free opacity of atomic carbon dominates the
continuum opacity from 1000 to 4000\,\AA, where around 80\,percent of the
integrated flux is emitted. We therefore chose carbon for the denominator of
abundance ratios, reducing the degree of correlation between abundances. Had
oxygen been used for this purpose, varying \logXY{C}{O}\ with other abundances
held constant relative to oxygen causes the line strengths of unrelated metals
to also vary (whereas varying \logXC{O} with other abundances held constant
relative to carbon does not have such an effect). For redder wavelengths, the
atomic carbon and atomic oxygen bound-free opacities are roughly equal up to
$\approx 6000$\,\AA, after which point the oxygen photoionisation opacity
dominates. Beyond 9000\,\AA\, metal free-free opacities overtake the metal
bound-free opacities.

While the best fitting model from this process showed generally good agreement
with the data (Figure~\ref{fig:GTCspec}, red), the formal uncertainties were
unrealistically small, i.e. 50\,K in $\Teff$, and a few 0.01\,dex in all other
parameters. To better estimate the uncertainties, we constructed 1-dimensional
grids of models for each parameter around the best fitting values. For $\Teff$
this meant models spanning $-1000$ to $+1000$ in steps of $50$\,K around the
best fitting value, and for $\log g$ and chemical abundances $1$\,dex above and
below the best fitting value in steps of $0.05$\,dex. These grids were visually
compared against the data to determine our quoted uncertainties. For the
surface gravity this was more complex than for normal white dwarfs, where
typically the $\log g$ is inferred from the pressure broadening of the hydrogen
or helium lines. Instead we found that increasing the $\log g$ changed the
strengths of many different metal lines from a variety of elements
simultaneously, with some becoming stronger and others weaker, leading to a
higher uncertainty than for normal white dwarfs.

We used a similar approach to determine upper-limits for hydrogen and helium,
constructing grids spanning many orders of magnitude in steps of $0.5$\,dex. We
then used the Bayesian approach developed by \citet{hollandsetal20-1} (i.e.
using a Jeffreys prior, $P(Z) \propto 10^{Z/2}$, where $Z = \logXC{H(e)}$) to
determine the 99th percentile upper-limits of $\logXC{H} < -4.1$\,dex and
$\logXC{He} < -1.5$\,dex respectively. As nickel is an important element
synthesised in type Ia supernovae, we used the same approach to determine an
upper-limit of $\logXC{Ni} < -2.9$\,dex focussing on a strong (unresolved)
\ion{Ni}{i} doublet at 5082\,\AA. We note that stronger \Ion{Ni}{i} lines are
expected at wavelengths between 3400--3600\,\AA\, (bluer than is covered by our
GTC spectrum), which may permit an improved upper-limit or measurement in the
future.

Evidence for rotation has been observed for supernova bound remnants in the
past for both the \lp\ \citep{hermesetal21-1} and \Dsix\
\citep{chandraetal22-1} classes of stars. In both cases, rotational periods on
the order of $\sim 10$\,hours are inferred from time-series photometry. While
we do not currently have time-series photometry available for \WD, it is still
possible to place constraints on rotation via broadening of spectral lines. We
found no evidence for rotational broadening of the spectrum, with an inferred
upper-limit of $v\sin i < 40$\,\kms. Indeed for a 10\,hr rotation period and
radius of 0.1\,\Rsun\ (which is close to the actual radius inferred in later
sections), the 14\,\kms\ equatorial rotation speed would be well below our
detection threshold for our $R=2500$ spectra.

Our best fitting spectroscopic parameters and their uncertainties (or
upper-limits) are presented in Table~\ref{tab:spec}. The best fitting model is
shown in all panels of Figure~\ref{fig:GTCspec} demonstrating good agreement in
the majority of lines. We note that a few of the strongest oxygen lines are
slightly too broad (such as the 5331\,\AA\ line), whereas others (such as the
4970\,\AA\ line) are slightly too narrow, suggesting that improved broadening
constants may be needed for more accurate modelling. \WD\ also shows a strong
feature of \Ion{C}{i} near 4470\,\AA\ which appears only weakly in our model.
This line blend has been observed in other carbon-rich atmospheres
\citep{hollandsetal20-1,kilicetal24-1}, showing a similar discrepancy,
suggesting improvements to the oscillator strengths of these lines are
required. Similarly, absorption is also missing from our model near the
6107\,\AA\ \Ion{O}{i} line and 5403\,\AA\ \Ion{Mg}{ii} line blend, indicating
some additional minor opacities are required (which may or may not be related
to the partially overlapping lines mentioned above). We have no reason to
believe these missing or underestimated lines will affect our estimated
atmospheric parameters.

\begin{table}
    \centering
    \begin{tabular}{l|c}
        \hline
        Parameter & value \\
\hline
\Teff      &  $15\,680\pm250$\,K \\
$\log g$   & $6.3\pm0.3$\,dex(cm\,s$^{-2}$) \\
$v_r$      & $384\pm6$\,\kms \\
$v\sin i$  & $<40$\,\kms \\
H  & $<-4.1$\,dex \\
He & $<-2.0$\,dex \\
O  & $+0.27\pm0.10$\,dex \\
Ne & $-1.86\pm0.25$\,dex \\
Mg & $-2.33\pm0.15$\,dex \\
Al & $-4.00\pm0.30$\,dex \\
Si & $-1.76\pm0.10$\,dex \\
S  & $-2.84\pm0.30$\,dex \\
Ca & $-2.85\pm0.10$\,dex \\
Fe & $-3.00\pm0.20$\,dex \\
Ni & $<-2.9$\,dex \\
\hline
    \end{tabular}
    \caption{Atmospheric parameters for \WD\ determined from our spectroscopic
    fitting. Abundances are logarithmic number ratios relative to carbon.}
    \label{tab:spec}
\end{table}

\subsection{Chemical abundance interpretation}

\begin{figure}
    \centering
    \includegraphics[width=\columnwidth]{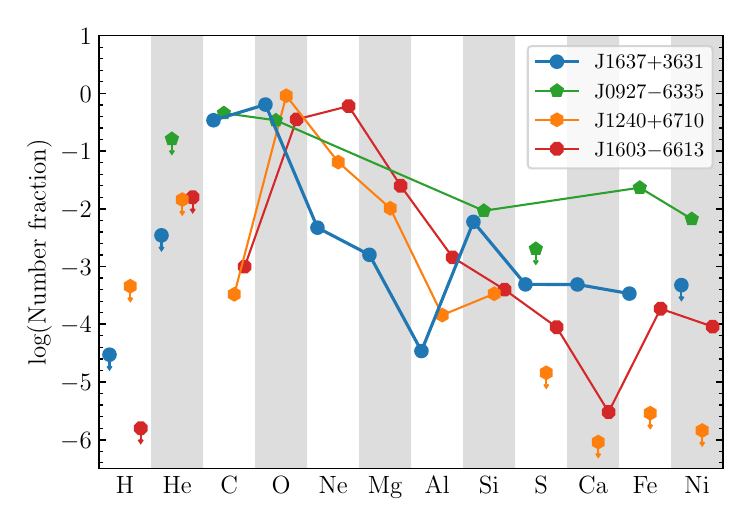}
    \includegraphics[width=\columnwidth]{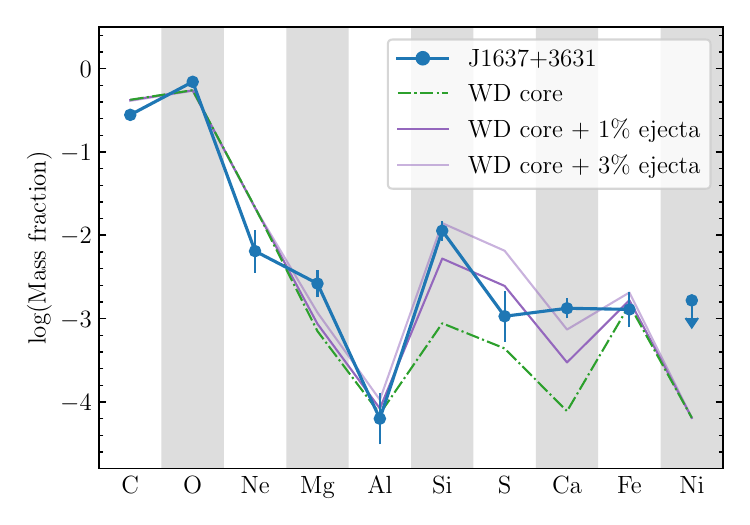}
    \caption{\emph{Top:} Atmospheric number fractions for \WD\ (blue circles)
    compared with the hot \Dsix\ star, J0927$-$6335 (green pentagons); the
    oxygen-rich white dwarf, SDSS\,J1240$+$6710 (orange hexagons); and one star
    in the \lp\ class, J1603$-$6613 (red octagons). \emph{Bottom:} Atmospheric
    mass fractions of \WD\ compared with the predicted core composition for a
    0.57\,\Msun\ white dwarf (green dot dash), and the same model but with 1
    and 3\,per\,cent mass added from D6 ejecta.
    }
    \label{fig:abundances}
\end{figure}

The abundances in Table~\ref{tab:spec} reveal an object dominated by carbon and
oxygen in a roughly 1:2 ratio. The atmosphere also contains moderate traces of
other light elements, i.e. Ne, Mg, and Al, as well as heavier elements, i.e.
Si, S, Ca, and Fe. These atmospheric abundances are shown in
Figure~\ref{fig:abundances} (top), normalised to the total abundance (by
number), and compared with other chemically peculiar white dwarfs. J0927$-$6335
is the hot \Dsix\ star discovered by \citep{el-badryetal23-1} and analysed by
\citep{werneretal24-2} from its UV spectrum (their results shown here).
SDSS\,J1240$+$6710 is the oxygen-rich white dwarf first identified by
\citet{kepleretal16-1}, and with a more detailed spectroscopic analysis
performed by \citet{gaensickeetal20-1} (their results shown here). J1603$-$6613
is one of the \lp-stars analysed by \citet{raddietal19-1}, and has the most
elements detected for any \lp-star so far. The three \lp-stars analysed by
\citep{raddietal19-1} have remarkably consistent abundances, and so we take
J1603$-$6613 as a representative member of the \lp-class. Neither hydrogen nor
helium are detected in any of the four objects, with upper-limits ruling out
these elements as major atmospheric constituents (a hydrogen upper-limit is not
provided for J0927$-$6335, though at its high \Teff, a non-detection rules out
hydrogen as a major atmospheric component, \citealt{werneretal24-1}), with
particularly good limits measured for \WD.

While all four stars share oxygen as a major atmospheric constituent, only \WD\
and J0927$-$6635 also have carbon comprising a large fraction of their
atmospheres, indicating that \WD\ is unrelated to the other two stars.
Furthermore, \WD\ is depleted in neon, magnesium, and aluminium compared to
SDSS\,J1240$+$6710 and J1603$-$6613, but instead shows higher abundances for
heavier elements, with a particularly striking spike for silicon, representing
around 1\,percent of the atmosphere, comparable to that measured for
J0927$-$6635. However, \WD\ and J0927$-$6335 differ by around 2\,dex for iron,
and by at least 1.2\,dex for nickel, though at the much higher \Teff\ of
J0927$-$6335, radiative levitation may be responsible for the atmospheric
enhancement of these abundances \citep{werneretal24-2}.

As well as similarities to J0927$-$6335, the C+O dominated atmosphere of \WD\
is superficially similar to the carbon and oxygen rich spectra reported for the
prototype \Dsix\ stars \citep{shenetal18-1}. \WD\ may therefore be a surviving
\Dsix\ remnant, where the present day abundances reflect the former donor white
dwarf core composition (with the outer envelope stripped by supernova shocks),
and then enhanced in heavier elements deposited from the supernova ejecta.
Using a MESA model for a 3\,\Msun\ star evolved up to the formation of a
carbon-oxygen core (see \citet{shen25-1} for further details), we took the
central composition as representative of the unaltered composition of a \Dsix\
donor which \citet{shen25-1} have suggested are fully convective (and therefore
chemically homogenous) stars. These data (Figure~\ref{fig:abundances}, bottom)
show generally consistent agreement from carbon up to aluminium, but also for
Fe. For intermediate mass elements (IMEs) -- specifically, silicon, sulphur,
and calcium -- our observations show clear enhancement compared to the expected
white dwarf CO core composition. These IMEs are also expected to be highly
abundant in the ejecta of \Dsix\ supernovae, particularly for velocities
between 10\,000 to 20\,000\,\kms\ \citep{boosetal21-1}. Importantly however,
the low velocity material is dominated by radioactive nickel, which is not seen
in the form of an Fe-excess.

We considered a simple model for the abundances of \WD, where the CO core of
the donor was enhanced with supernova ejecta, with elements impacting the donor
in reverse order of the ejecta velocity, i.e. where the fastest ejecta becomes
buried by each successive layer. Since the nickel-rich outer most layers will
be only marginally bound to the inflated donor, radioactive decay powered winds
\citep{shen+schwab17-1,bhatetal25-1} are then assumed to remove most of this
element, with any remainder decaying to iron. Finally, as \WD\ evolves to a
fully convective star, these ejecta abundances are mixed homogenously within
the initial core material. Of course this model is only an approximation, as it
does not consider time-dependent mixing within the donor envelope, nor does it
include detailed calculations for the fraction of each element removed via
$^{56}$Ni decay driven winds.

\citet{boosetal21-1} performed 2D simulations of white dwarf double detonations
and their ejecta, with a variety of white dwarf masses and helium shell
thicknesses. We took their 1\,\Msun\ model (for the exploding primary) using
the thinnest helium shell value (specifically their
\texttt{2d\_decayed\_ejecta\_t35e7\_d2e5\_m100\_v500} data), and for simplicity
assumed equatorial ejection abundances. The abundances are provided as mass
fractions as a function of velocity, $f_Z(v)$, and with the ejecta density,
$\rho(v)$, also given. The fastest material has low density and should
therefore contribute very little to the accreted total. On the other hand, the
slowest material has high density, but as stated previously, is nickel-rich.
Since we do not observe significant iron-enrichment, we therefore assumed that
material with nickel fractions above 10\,percent were removed via decay powered
winds, corresponding to a minimum velocity of 11\,000\,\kms. For a given
element, $Z$, the integrated total is then given by
\begin{equation}
I_Z = \int_{11}^{45} f_Z(v) \rho(v) \,\dd v,
\end{equation}
where $v$ is in units of 1000\,\kms\ and the data by \citet{boosetal21-1} are
calculated up to 45\,000\,\kms. Of course the magnitude and units of these
integrals are arbitrary, and would require the donor cross-sectional area and
accretion timescale, to directly calculate the integrated mass. However as we
are specifically interested in their ratios, the mass fractions of the accreted
ejecta, $F_Z$, are given by $F_Z = I_Z / \sum_k I_k$.

To determine the present day atmospheric abundances, we combined the core and
ejecta abundances according to $F_{Z,\mathrm{atm}} = (1-q)F_{Z,\mathrm{core}} +
q F_{Z,\mathrm{ejecta}}$, where $q$ is the ejecta mass-fraction of the combined
core and ejecta material. Our results are shown in Figure~\ref{fig:abundances}
(bottom) for values of $q=0.01$ and $q=0.03$. For 1\,percent accreted ejecta, a
mediocre compromise is is seen for silicon, sulphur, and calcium compared to
the observed abundances of \WD. For 3\,percent ejecta, good agreement is seen
between for both silicon and calcium. However, the sulphur abundance is
over-predicted by 0.7\,dex, more than 2$\sigma$ above the measurement.

One possible explanation is that systematic uncertainty in the modelling has
lead to this (weakly detected) element to be underestimated. In
Figure~\ref{fig:GTCspec}, the \Ion{S}{ii} lines near 5650\,\AA\ are well
modelled, whereas the \Ion{S}{i} lines near 6750\,\AA\ \Ion{S}{i} are clearly
underestimated in strength. Increasing the atmospheric sulphur abundance by
0.7\,dex provides a reasonable estimate for the strength of those \Ion{S}{i}
lines without any substantial over-prediction of the 5650\,\AA\ \Ion{S}{ii}
equivalent widths. However, a +0.7\,dex abundance also results in additional
sulphur lines appearing in our model which are not observed in the data, such
as \Ion{S}{ii} lines at 4164\,\AA\ and 5214\,\AA. Therefore the abundance
provided in Table~\ref{tab:spec} provides a best compromise from our fitting
procedure.

The quality of the atomic data in NIST \citep{kramidaetal24-1} is rated C+
($\leq 18$\,percent accuracy) for the 5650\,\AA\ \Ion{S}{ii} lines, whereas the
the 6750\,\AA\ lines are rated D+ ($\leq 40$\,percent accuracy). The other
strong sulphur lines covered by our data have accuracy ratings ranging from C+
to E ($>50$\,percent accuracy). Alternatively, the yields calculated by
\citet{boosetal21-1} may simply over-predict the amount of sulphur produced by
double-detonations. Either way, improved oscillator strengths for optical
\Ion{S}{i--ii} transitions will help to resolve this discrepancy.

While we have shown results from one specific dataset in
Figure~\ref{fig:abundances}, \citet{boosetal21-1} provide a variety of data for
calculations over a range of primary white dwarf masses, temperatures, and
helium shell thicknesses. Experimenting with the other provided yield data, we
found that the $F_Z$ ratios are fairly consistent across all calculations, with
$F_\mathrm{S}/F_\mathrm{Si}$ between $0.46$--$0.50$, and
$F_\mathrm{Ca}/F_\mathrm{Si}$ between $0.05$--$0.07$.

Our assumption that nickel is removed from the donor by radiation powered winds
was motivated by the relative absence of iron, which would otherwise be
expected to dominate the accreted ejecta abundances. This led to the conclusion
that \WD\ is enhanced in IMEs, in apparent contrast to the results of
\citet{wongetal24-1} who instead found the ejecta to be dominated by iron group
elements (IGEs). However, their short timescale hydrodynamical simulations only
cover the initial deposition of the IGE-dominated ejecta onto the donor, where
the longer term evolution and radioactive decay of the accreted material is not
considered. Their post-explosion evolutionary models using MESA (Modules for
Experiments in Stellar Astrophysics) do not account for the accreted ejecta.

In conclusion, we find that \WD\ is \emph{chemically} consistent as the donor
of a \Dsix\ supernova, where the observed abundances reflect the stripped core
of a CO white dwarf, enhanced in Si- and Ca-rich supernova ejecta by
1--3\,percent. Though we note some possible inconsistencies in the sulphur
abundance and require the lowest velocity nickel-rich ejecta to be removed via
radioactive decay powered winds.

\section{Confirming the \texorpdfstring{\Dsix}{D6} nature of
\texorpdfstring{\WD}{J1637+3631}}
\label{sec:confD6}

In addition to the exotic atmospheric composition, the other key signature of
\Dsix\ remnants is their extreme kinematics with Galactic space motions
exceeding 1000\,\kms, resulting from their ejections from tight,
double-degenerate binary systems. The radial velocity component of \WD\ is only
around 400\,\kms\ (Table~\ref{tab:spec}), and is therefore insufficient to
fully establish the \Dsix\ nature. The tangential velocity, $v_\perp$, is
instead required to confirm this scenerio.

\WD\ has a moderate proper-motion at $|\mu|=67.08\pm0.56$\,mas\,yr$^{-1}$ (see
Table~\ref{tab:astromphot}), or equivalently $318\pm3$\,\kms\,kpc$^{-1}$.
Ideally, the distance would be inferred from the \emph{Gaia} parallax. As of
\emph{Gaia} DR3, however, the imprecise parallax of $0.3319\pm0.4776$\,mas is
only sufficient to claim that \WD\ is at least $\sim 0.5$\,kpc from the Sun (at
the 3$\sigma$ level), and so space-motions ranging from only a few hundred to
many thousands of \kms\ have near-equal likelihood, and are therefore
kinematically consistent with a variety of explosion scenarios. Additional
information is therefore needed to constrain the current distance and $v_\perp$
of \WD. Fortunately the spectroscopic $\log g$ serves this purpose, as it links
the mass of \WD, $M_\star$, to its radius, $R_\star$, which in turn constrains
the distance, $D$, via the observed photometric fluxes and spectral model.
Essentially, for some fixed solid-angle for the stellar disc, $M_\star \propto
g D^2$. Therefore, if we require that $M_\star < 1.4$\,\Msun, and $g$ is
constrained by a measurement, an upper-bound can be placed on $D$. If a more
informed prior on $M_\star$ can be assumed, then the distances consistent with
the data can be further refined.

To this end, we developed a simple Bayesian model to estimate the distance,
mass, and radius of \WD. The likelihood was constructed from the \emph{Gaia}
DR3 parallax, spectroscopic $\log g$, and photometry
(Table~\ref{tab:astromphot}). Each photometric point was assumed to have a
0.01\,mag systematic uncertainty in its absolute calibration in addition to the
statistical errors, the latter of which is particularly small for the
\emph{Gaia} $G$-band. Additionally we included the \Teff\ and the interstellar
reddening, \ebv, as nuisance parameters. For \Teff, we simply interpolated
between our 1D grid of models which we previously used to estimate the $\Teff$
uncertainty. For \ebv, we used the extinction model of \citet{gordonetal23-1},
which we applied to our spectral model before calculating synthetic photometry
by integrating over the appropriate pass bands. Each free parameter also
requires careful consideration for its prior distribution. For the distance we
assumed $P(D) \propto D^2$, reflecting a constant space density along the line
of sight, which is a reasonable approximation for detectable objects close to
the Galactic plane.\footnotemark\ For the mass, we assume that \WD\ was a
normal CO core white dwarf (as evidenced by its spectrum) that lost some
unknown fraction of its mass (e.g. due to donating mass to a companion, or from
being stripped by supernova ejecta). We therefore adopted a prior distribution
taking a two-component Gaussian mixture model for the white dwarf mass
distribution (given by \citealt{obrienetal24-1}), and convolved with a
uniformly distributed fractional mass loss, i.e. replacing the Gaussian
components with the corresponding Gaussian integrals. For the stellar radius,
we assumed a log-uniform distribution between $10^{-3}$ and $1$\,\Rsun. For the
\Teff\ parameter, we used a Gaussian prior from the spectroscopic measurement,
and finally for the reddening we used the Poisson Jeffreys prior of $P(\ebv)
\propto \ebv^{-1/2}$.

\footnotetext{Beyond 10s of kpc, the Milky Way effectively becomes a point
source for ejecting bound supernova remnants, at which point the density should
drop off as $D^{-2}$. While this will result in a flat distance prior by
itself, the limiting magnitude of \emph{Gaia} will cause the distance
distribution for \emph{observable} \Dsix\ stars to eventually drop to zero.
This is discussed further in Section~\ref{sec:future}.}

To sample the posterior distribution, we used the \textsc{python} package
\textsc{pocoMC} \citep{karamanisetal22-1, karamanisetal22-2}, which uses a
preconditioned Monte-Carlo (PMC) algorithm. This algorithm draws a specified
number of particles from the prior distribution, $P(\theta)$, evolving the
ensemble towards the posterior distribution $P(\theta|\mathbf{x})$ by gradually
increasing an inverse-temperature, $\beta$, from 0 to 1 according to 
\begin{equation}
    P(\theta|\mathbf{x};\beta) \propto P(\theta) L(\mathbf{x}|\theta)^\beta,
\end{equation}
where $L(\mathbf{x}|\theta)$ is the likelihood of the data $\mathbf{x}$ with
parameters $\theta$. Furthermore, preconditioning is achieved at each step
using a normalising flow, which allows the target distribution to be
transformed to a simpler one that is easier to sample. Compared to other
sampling methods such as Markov-Chain Monte-Carlo, the \textsc{pocoMC}
implementation of PMC offers rapid convergence, while still performing well in
complex (potentially multi-modal) parameter spaces, and with large numbers of
free-parameters (a feature which is less relevant here, but which we exploit in
the following section). Furthermore, \textsc{pocoMC} only requires
prescriptions of the prior distribution and likelihood as inputs, without the
need to supply analytical gradients (as in the case of techniques such as
Hamiltonian Monte-Carlo). The online documentation\footnotemark\ and
accompanying references (above) should be consulted for a more in depth
description of the algorithm.

\footnotetext{\url{https://github.com/minaskar/pocomc}\\
\url{https://pocomc.readthedocs.io/en/latest/}}

\begin{figure} \centering
\includegraphics[width=\columnwidth]{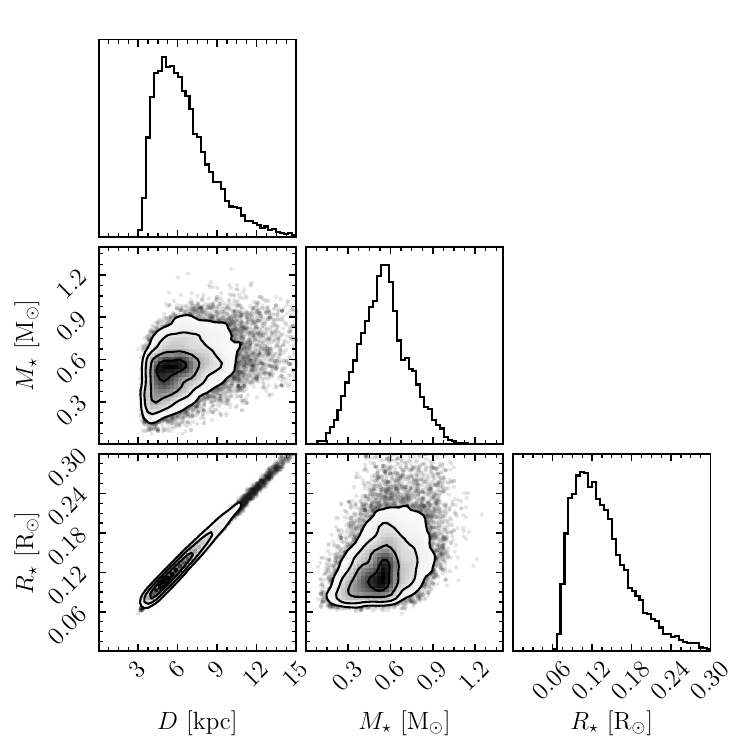}
\caption{Corner plot to our fit to the distance, mass, and radius of \WD, using
the parallax, $\log g$, and photometry as constraints.}
\label{fig:corner_simple}
\end{figure}

We sampled our model using the default \textsc{pocoMC} parameters (as of
version 1.2.6). Specifically \texttt{n\_effective=512}, \texttt{n\_active=256},
which set the number of effective particles and active particles at each step,
and also \texttt{flow='nsf6'} (neural spline flow with 6 transformations) for
the choice or normalising-flow model. After convergence to $\beta=1$, the total
number of particles was allowed to grow to 16\,384 to provide a large number of
samples from the posterior distribution. The corner plot for our parameters of
interest, $D$, $M_\star$ and $R_\star$ are shown in
Figure~\ref{fig:corner_simple}, where $D=6.1_{-1.6}^{+2.6}$\,kpc,
$M_\star=0.55_{-0.17}^{+0.19}$\,\Msun, and
$R_\star=0.128_{-0.034}^{+0.054}$\,\Rsun. For the nuisance parameters, \Teff\
simply reflects the Gaussian prior, and has only a small correlation with the
other nuisance parameter, \ebv, which was found to have a value of
$0.086\pm0.013$.

As hoped, the inclusion of $\log g$ in the likelihood and some simple
assumptions about the range of feasible masses for \WD\ provides reasonable
constraints on all three parameters. In particular, the distance is measured to
approximately $3\sigma$ precision. Scaling by the proper-motion of
$318\pm3$\,\kms\,kpc$^{-1}$ therefore yields $v_\perp =
1950_{-530}^{+810}+$\,\kms. Combining this with the radial velocity gives a
speed of $1980_{-510}^{+790}$\,\kms\ in the heliocentric reference frame.
Transforming these samples to the Galactocentric frame gives a Galactic motion
of $2070_{-490}^{+790}$\,\kms. Therefore, in addition to the C+O composition
enriched with nuclear burning products, this extreme velocity essentially
confirms that \WD\ is indeed the ejected survivor of a \Dsix\ explosion.

\section{Kinematic analysis}
\label{sec:kin}

In Section~\ref{sec:confD6}, we presented a Bayesian model incorporating the
\emph{Gaia} parallax, spectroscopic $\log g$, and photometry to constrain the
distance, mass, radius, and interstellar reddening. This was in turn used to
infer the Galactocentric velocity of \WD\ corroborating our earlier assertion
that it was ejected from a \Dsix\ supernova. With the distance and velocity
samples in hand, it would be a relatively simple task to trace-back orbits to
the point they cross the Galactic plane, thereby finding an approximate
location of the explosion, and time of flight (assuming \WD\ to be ejected from
the mid-plane of the disc). Such simulations have been performed for other
runaway stars in the past
\citep{geieretal15-1,raddietal18-2,irrgangetal18-1,shenetal18-1,raddietal19-1,
ruffinietal19-1,evansetal20-1,el-badryetal23-1}. However, by constraining the
distance first and \emph{then} constraining the ejection site as two separate
steps, we lose any potential correlations between parameters of those separate
calculations -- or in other words, more precise parameters may be found by
attempting to constrain the present distance and the ejection-site
simultaneously. We also consider that the ejection-site should not be
restricted to the Galactic mid-plane, and could instead occur above or below
the plane, approximately proportional to the stellar density in the disc.

To that end, we extended our Bayesian model from before, incorporating Galactic
dynamics in order to trace back \WD\ to its potential ejection site. We used
the \textsc{python} package \textsc{galpy} \citep{bovyetal15-1}, which allows
orbital integration using a variety of potentials including for within the
Milky Way galaxy. We used the included \texttt{MWPotential2014} potential in
our calculations which consists of three components in order to model the disc,
bulge, and halo. The disc is modelled with a Miyamoto-Nagai potential
\citep{miyamoto+nagai75-1}, the bulge is modelled with a power law density
profile with exponential cutoff \citep{bovyetal15-1}, and the halo uses a
Navarro-Frenk-White potential \citep{navarroetal97-1}. While we use
\textsc{astropy} for working with sky coordinates and conversions to the
Galactocentric frame \citep{astropy13-1,astropy18-1,astropy22-1}, internally
\textsc{galpy} works in a units system of natural units based on the distance
to the Galactic centre, $R$, and the Sun's Galactic orbital speed $v$. When
converting to physical units, by default \textsc{galpy} uses $R=8$\,kpc, and
$v=220$\,\kms. To improve compatibility with the \textsc{astropy}
Galactocentric frame, we instead used $R=8.122$\,kpc \citep{gravitycollab18-1}
for all conversions, taking care that this did not lead to any inconsistencies.
Note that where \textsc{astropy} uses a right-handed coordinate system, a left
handed system is used by \textsc{galpy}, and so further care was taken to
ensure correctness throughout.

In addition to the five free parameters from before (Section~\ref{sec:confD6}),
we included the three Galactocentric velocity components for \WD\ at its
current location, $\mathbf{v} = (v_x, v_y, v_z)$, which were each given flat
priors from $-3000$\,\kms\ to $+3000$\,\kms. We also included a time of flight
parameter, $t$, for integrating the past orbit of \WD. From the distance
posterior in Section~\ref{sec:confD6} and Monte-Carlo sampling the
proper-motion, we infer a value of $Z/v_z = 4.5\pm0.4$\,Myr (where $Z$ is the
height above the Galactic plane in Galactocentric frame), which approximates
the travel time from the mid-plane to the current location of \WD. We therefore
adopted a uniform prior from 0 to 10\,Myr on $t$. Finally, a fractional
parameter, $f$ (with a uniform prior from 0 to 1), was also introduced to
represent the mass fraction of the donor that survives the supernova, where the
mass free-parameter from before now represents the donor mass at the time of
detonation (denoted $M_\mathrm{don}$), and where the current mass of \WD\ can
be calculated as $M_\star=fM_\mathrm{don}$. The need for this extra free
parameter, $f$, will become apparent in a second simulation including tighter
physical constraints. The priors on all parameters appearing in
Section~\ref{sec:confD6} remain unchanged.

Compared to before, the likelihood, $L$, is more complex, incorporating five
independent components,
\begin{equation}
    L = L_\mathrm{astrom} \times L_{v_r} \times L_{\log g} \times L_\mathrm{phot} \times \rho_\mathrm{ej},
\end{equation}
with each term detailed below. Firstly, we introduce the \emph{Gaia}
astrometric likelihood $L_\mathrm{astrom}(\varpi, \mu | D, \mathbf{v};
\Sigma)$, where $\varpi$ is the parallax (corrected for the \emph{Gaia} DR3
zero-point), $\mu$ is the proper motion (both components), and $\Sigma$ is
their covariance matrix including the off-diagonal terms provided by
\emph{Gaia}. $L_\mathrm{astrom}$ is simply a multivariate-normal distribution
where \textsc{astropy} was used to perform coordinate conversion from distance
and Galactocentric velocity to the estimated parallax and proper-motion to be
compared with the measured values. The velocity to proper-motion conversion
implicitly requires the right ascension and declination, though these were
treated as fixed-values due to their sub-mas precision.

The spectroscopic radial-velocity, $v_r$ has a simple Gaussian likelihood
written as $L_{v_r}(v_r|\mathbf{v}, M_\mathrm{don}, f, R; \sigma_{v_r})$. The
conversion from Galactocentric velocity components to inferred radial-velocity
was performed as part of the previously described coordinate conversion.
However, since the $v_r$ measurement is independent of the other astrometry, we
treat it as a separate likelihood term. We note that $L_{v_r}$ also depends on
$M_\mathrm{don}$, $f$ and $R_\star$, as the spectroscopic measurement includes
both the relative motion and gravitational redshift. Whilst the values of
$M_\star$ and $R_\star$ found in Section~\ref{sec:confD6} imply a gravitational
redshift of only a few \kms, this is comparable to our measurement uncertainty,
and so should not be neglected.

The likelihood for the $\log g$ is similar to before, i.e. a Gaussian
likelihood, but now $M_\star = fM_\mathrm{don}$, and so this term now depends
on three free-parameters rather than two. The photometric term in the
likelihood remains unchanged, as it depends only on $D$, $R_\star$, \Teff, and
\ebv, which were parameters in Section~\ref{sec:confD6}.

At this point, for a given $D$ and $\mathbf{v}$, the model has no way to
constrain $t$. The ejection site, $\mathbf{r}_0 = (X_0, Y_0, Z_0)$, is of
course a function of $D$, $\mathbf{v}$ and $t$, as they are the inputs to the
orbital integration. However, we also wished to impose that the ejection site
is more likely to be located in Galactic regions with higher stellar density.
We therefore weighted the likelihood proportional to the mass-density in the
Milky Way, i.e. $\rho_\mathrm{ej}(D, \mathbf{v}, t) = \rho(\mathbf{r}_0(D,
\mathbf{v}, t))$. While the mass-density within the Galaxy, and number density
of binaries that could produce \Dsix\ detonations need not be strictly
proportional, the mass density is the more accessible quantity, and within the
Galactic disc are likely to be similar. We calculated the mass density directly
from the \texttt{MWPotential2014} potential model within \textsc{galpy}, where
the library provides an analytic form for the potential components (without
resorting to numerically calculating Poisson's equation). However, we only used
the bulge and disc components of the potential, ignoring the dark-matter
dominated halo component. Formally $\rho_\mathrm{ej}$ is not a likelihood as
the parameters do not define a distribution from which some data might be
drawn. Instead it constitutes part of a joint prior on $D$, $\mathbf{v}$, and
$t$, that restrict their possible combinations to those that originate from
dense regions of the Galaxy. However, this joint prior is not trivial to sample
in terms of $D$, $\mathrm{v}$, and $t$, which is required for the
initialisation of our fitting approach. We therefore combined this with the
likelihood terms for technical simplicity.

\subsection{Results and ejection site location}

\begin{figure*}
    \centering
    \includegraphics[width=\textwidth]{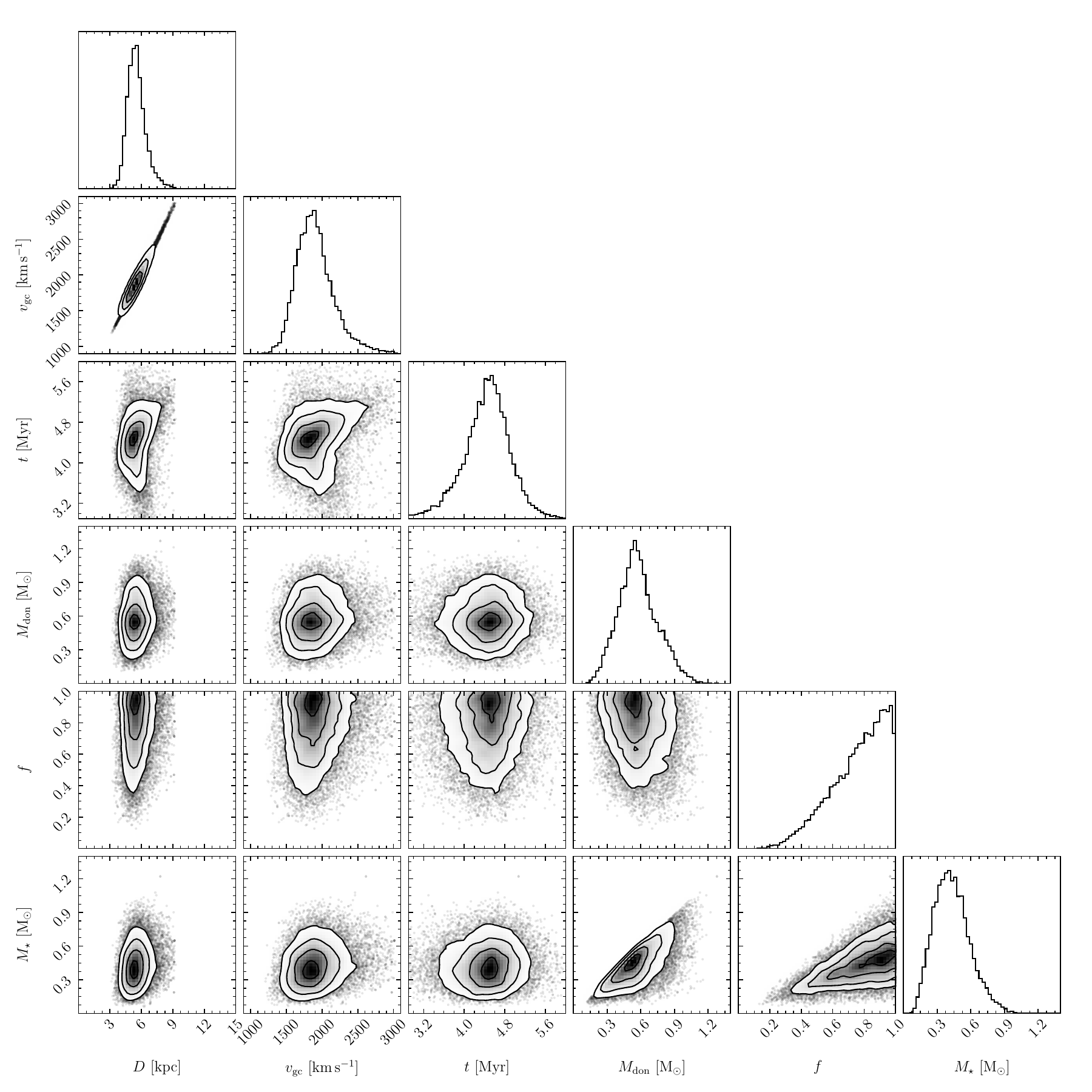}
    \caption{Corner plot for a subset of parameters in Fit 1. The derived
    parameter $M_\star$ is also included, where $M_\star = fM_\mathrm{don}$.
    Similarly $v_\mathrm{gc}$ is determined from the modulus of the individual
    components of $\mathbf{v}$. The distance, $D$, is shown over the same range
    as in Figure~\ref{fig:corner_simple}, demonstrating the substantial
    improvement in precision. $M_\mathrm{don}$ and $M_\star$ are also shown
    over the same range as $M_\star$ in Figure~\ref{fig:corner_simple}.
    } 
    \label{fig:corner_traceback}
\end{figure*}

As before, we used \textsc{pocoMC} to sample the posterior
distribution,\footnotemark\ and using the same sampler parameters, and same
final number of particles in the distribution. The best fitting values are
given under `Fit 1' in Table~\ref{tab:kinematics} (with the absence of \Teff,
which once more simply reflects the Gaussian prior). Table~\ref{tab:kinematics}
also includes other quantities derived from our fitted parameters: $X_0$,
$Y_0$, and $Z_0$ are the coordinates inferred for the ejection site in
Galactocentric coordinates, and $\rho_{XY}, \rho_{XZ}, \rho_{YZ}$ are their
correlation coefficients. Additionally, $v_\mathrm{gc} = |\mathbf{v}|$ is the
Galactocentric speed of \WD\ at the current epoch, and $v_\mathrm{ej}$ is the
ejection speed of \WD\ after subtracting the component of Galactic orbital
velocity at the ejection site. Finally, $M_\star$ is simply $f M_\mathrm{don}$.
A corner plot for $D$, $v_\mathrm{gc}$, $t$, $M_\mathrm{don}$, $f$, and
$M_\star$ is shown in Figure~\ref{fig:corner_traceback}.

\footnotetext{In an earlier iteration of this model we had attempted to sample
the posterior using MCMC methods such as \textsc{emcee}
\citep{foremanmackeyetal13-1}, however we found that the high-dimensionality of
the problem caused slow convergence, with the distribution having barely burned
in after several days of computation. Instead \textsc{pocoMC} allowed the
posterior to be sampled in less than one hour on the same machine, permitting
faster development.}

\renewcommand{\arraystretch}{1.2}
\begin{table}
    \centering
    \begin{tabular}{l|ccc}
        \hline
        Parameter                & Fit 0                     & Fit 1                     & Fit 2                  \\
\hline        
$D$ [kpc]                & $6.1_{-1.6}^{+2.6}$       & $5.4_{-0.7}^{+0.9}$       & $5.1\pm0.4$            \\
$v_x$ [\kms]             &                           & $-1440_{-260}^{+220}$     & $-1350\pm120$          \\
$v_y$ [\kms]             &                           & $930_{-60}^{+70}$         & $910\pm30$             \\
$v_z$ [\kms]             &                           & $760_{-70}^{+80}$         & $740\pm40$             \\
$t$ [Myr]                &                           & $4.5_{-0.5}^{+0.4}$       & $4.4\pm0.4$            \\
$M_\mathrm{don}$ [\Msun] &                           & $0.56_{-0.16}^{+0.20}$    & $0.57_{-0.14}^{+0.18}$ \\
$f$                      &                           & $0.79_{-0.22}^{+0.15}$    & $0.78_{-0.22}^{+0.16}$ \\
$R_\star$ [\Rsun]        & $0.128_{-0.034}^{+0.054}$ & $0.114_{-0.016}^{+0.019}$ & $0.108\pm0.009$        \\
\ebv                     & $0.086\pm0.013$           & $0.086\pm0.013$           & $0.086\pm0.013$        \\
\hline               
$X_0$ [kpc]              &                           & $+0.4_{-1.4}^{+2.0}$      & $+0.0\pm0.9$           \\
$Y_0$ [kpc]              &                           & $-0.8_{-0.3}^{+0.4}$      & $-0.8\pm0.4$           \\
$Z_0$ [kpc]              &                           & $+0.1_{-0.3}^{+0.4}$      & $+0.1\pm0.3$           \\
$\rho_{XY}$              &                           & $-0.20$                   & $-0.49$                \\
$\rho_{XZ}$              &                           & $-0.17$                   & $-0.48$                \\
$\rho_{YZ}$              &                           & $+0.973$                  & $+0.972$               \\
$|\mathbf{r}_0|$ [kpc]   &                           & $1.4_{-0.6}^{+1.3}$       & $1.1_{-0.3}^{+0.6}$    \\
$v_\mathrm{gc}$ [\kms]   & $2070_{-490}^{+790}$      & $1870_{-220}^{+280}$      & $1790\pm120$           \\
$v_\mathrm{ej}$ [\kms]   &                           & $1870_{-300}^{+360}$      & $1750\pm160$           \\
$M_\star$ [\Msun]        & $0.55_{-0.17}^{+0.19}$    & $0.42_{-0.15}^{+0.17} $   & $0.41_{-0.14}^{+0.17}$ \\
$L_\star$ [\Lsun]        & $0.89_{-0.41}^{+0.91}$    & $0.70_{-0.18}^{+0.25} $   & $0.63_{-0.10}^{+0.11}$ \\
\hline
    \end{tabular}
    \caption{Results from our orbital fits to \WD. The explicitly fitted
    parameters are listed first. The second group refer to quantities derived
    from the fitted parameters. \Teff\ is not shown as the results simply
    reflect the measured prior, instead allowing more freedom in the other
    parameters such as \ebv. Parameters from Section~\ref{sec:confD6} are
    listed under Fit~0 (note that $M_\star$ is the fitted parameter here).
    Fit~1 represents our adopted parameters for \WD, where the results from
    Fit~2 are potentially subject to uncertainties in the donor $\Teff$/radius
    at the time of explosion.
    }
    \label{tab:kinematics}
\end{table}
\renewcommand{\arraystretch}{1.0}

Compared to the results in Figure~\ref{fig:corner_simple} (which are provided
as Fit 0 in Table~\ref{tab:kinematics}), the distance to \WD\ is refined to
$5.4_{-0.7}^{+0.9}$\,kpc.\footnotemark\ The improved precision on the distance
is evident by comparing the top-left panels of Figures~\ref{fig:corner_simple}
and \ref{fig:corner_traceback}, which cover the same range of $D$ values. Since
$v_\perp$  is directly proportional to the distance, the components of
$\mathbf{v}$ (and hence $v_\mathrm{gc}$) are strongly correlated with $D$. The
precision on the radius, $R_\star$, is improved by a similar amount to $D$, as
they are also highly correlated. In contrast \ebv\ shows no improvement as it
depends only on the $\Teff$ prior and the photometry which are unchanged.

\footnotetext{As an interesting aside, because the distances involved are on
the order of kpc, and the tangential velocity on the order 1000\,\kms, the
observed sky position significantly lags the true current location. At a
distance of $5.4_{-0.7}^{+0.9}$\,kpc, the light travel time is $18\pm3$\,kyr.
Therefore \WD\ will have moved $31_{-8}^{+11}$\,pc tangentially or
$20\pm3$\,arcmin since the observed light was emitted.}

\begin{figure*}
    \centering
    \includegraphics[width=\textwidth]{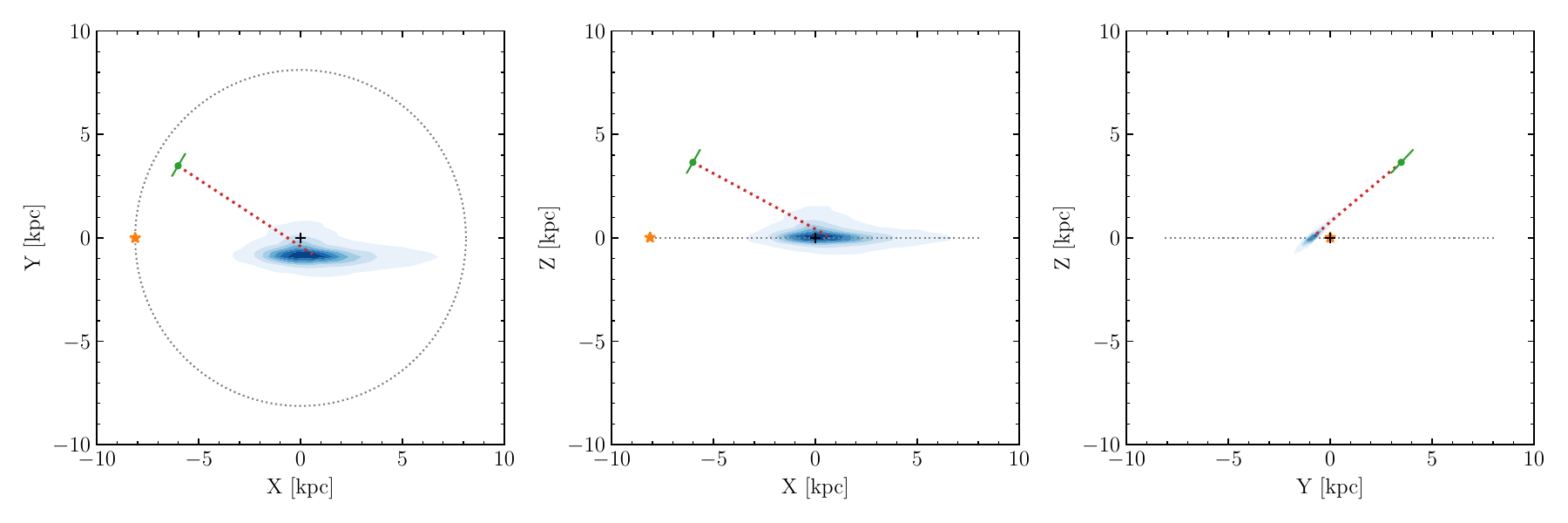}
    \caption{Ejection site and trajectory of \WD\ viewed along the three
    Cartesian axes, using posterior parameters from Fit 1. The current location
    of \WD\ and its distance uncertainty are shown by the green circle. The
    probability density for the ejection site is shown by the varying shades of
    blue. The trajectory for the median parameters is shown by the red dotted
    line. The black cross and orange star symbols indicate the location of the
    Galactic centre and Sun, respectively. The black dotted circle/lines
    correspond to the orbit of the Sun around the Milky Way.
    }
    \label{fig:birthsite1}
\end{figure*}

Figure~\ref{fig:birthsite1} shows the distribution of $X_0$, $Y_0$, and $Z_0$
within the Galaxy, as well the trajectory taken by \WD\ to its current
location. The distribution appears to overlap the Galactic centre when viewed
in the $X$-$Y$ and $X$-$Z$ planes. Viewed from the $Y$-$Z$ plane, however, it
is clear that \WD\ cannot have originated from the Galactic centre. In essence,
the posterior of $X_0$, $Y_0$, $Z_0$ is approximately an oblate Gaussian
ellipsoid, whose orientation avoids it overlapping the Galactic centre. The
volume enclosed by the $1\sigma$ Gaussian surface is $0.37$\,kpc$^3$,
calculated according to $\tfrac{4\pi}{3}|\Sigma_{XYZ}|^{1/2}$, where
$\Sigma_{XYZ}$ is the covariance matrix of $X_0, Y_0, Z_0$, and can be
constructed from the information provided in Table~\ref{tab:kinematics}. We
also calculated the probability that \WD\ originated from the disc versus the
bulge. For each sample, $i$, we calculated the disc and bulge densities
($\rho_\mathrm{disc}$ and $\rho_\mathrm{bulge}$, respectively) from the
\textsc{galpy} disc and bulge potentials. The probability that \WD\ was ejected
from the disc, $P_\mathrm{disc}$, is then given by
\begin{equation}
    P_\mathrm{disc} = \frac{1}{N}\sum_i^N
     \frac{\rho_\mathrm{disc}(\mathbf{r}_{0,i})}{\rho_\mathrm{disc}(\mathbf{r}_{0,i})+\rho_\mathrm{bulge}(\mathbf{r}_{0,i})},
\end{equation}
and similar for $P_\mathrm{bulge}$. We subsequently found that \WD\ has a
83\,percent probability of originating from the disc and a 17\,percent
probability of originating from the bulge.

\subsection{Comparison with other objects and evolutionary models}

\begin{figure}
    \centering
    \includegraphics[width=\columnwidth]{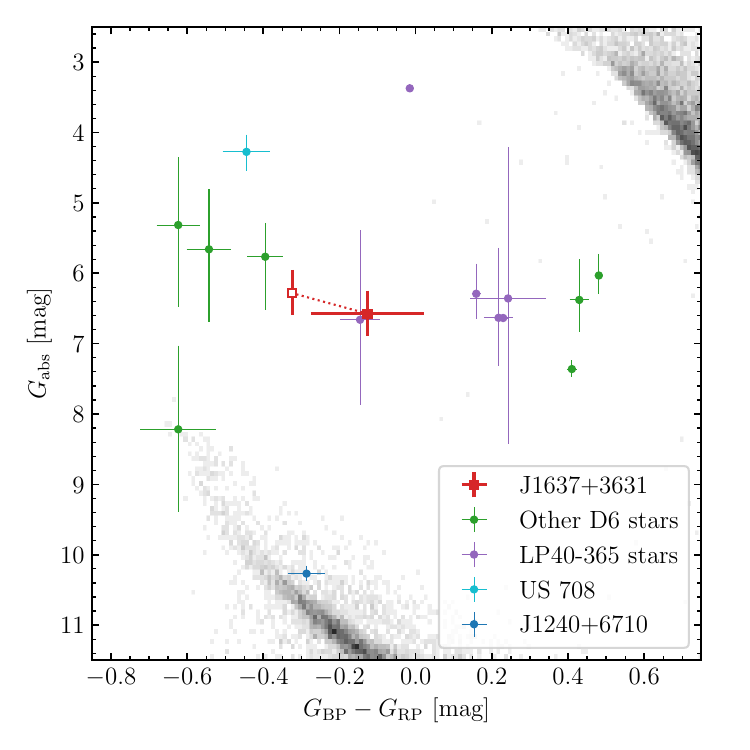}
    \caption{Hertzsprung-Russell diagram of runaway stars associated with
    supernovae. For \WD, the distance used is from our kinematic analysis (Fit
    1). The solid red square only uses the \emph{Gaia} photometry and inferred
    distance, whereas the hollow square uses synthetic photometry from our best
    fitting spectral model, and thus excludes interstellar reddening. Objects
    identified by \citet{el-badryetal23-1} have been de-reddened by the values
    given therein. The white dwarf and main sequences are shown in greyscale.}
    \label{fig:hrd}
\end{figure}

With our refined estimate of $D$, we are now able to place \WD\ on the
Hertzsprung-Russell diagram alongside other runaway white dwarfs and related
stars. This is shown in Figure~\ref{fig:hrd}, where the solid red square
corresponds to the \emph{Gaia} DR3 photometry as given. We also show the
location of \WD\ without reddening, with synthetic photometry calculated
directly from our best fitting spectral model. Compared to the previously
identified \Dsix\ stars, \WD\ sits between the redder/cooler objects found by
\citet{shenetal18-1} and the bluer/hotter objects published by
\citet{el-badryetal23-1}, reflecting its intermediate \Teff. Although
approximately double the \Teff\ of the cool \Dsix\ stars, \WD\ has a comparable
absolute magnitude, indicating that its 0.1\,\Rsun\ radius is smaller than
those objects.

\citet{shen25-1} recently calculated evolutionary tracks for the surviving
donors of \Dsix\ supernovae. In their Figure~1, the fully convective CO donors
initially contract almost isothermally at $\Teff \approx 6000$\,K, before
eventually becoming radiative and evolving towards higher $\Teff$ and briefly
towards higher luminosities. Eventually their cores become degenerate, and the
\Dsix\ stars return to the white dwarf cooling sequence. Our spectroscopic
$\Teff=15\,680\pm250$ combined with the inferred radius,
$R_\star=0.114_{-0.016}^{+0.019}$\,\Rsun, yields a luminosity of
$L_\star=0.70_{-0.18}^{+0.25}$\,\Lsun. These values imply a surviving donor
mass of 0.18--0.20\,\Msun\ (\citealt{shen25-1}, Figure~1) which is smaller than
our estimate of $0.42_{-0.15}^{+0.17}$\,\Msun, as given in
Table~\ref{tab:kinematics}. The cumulative distribution of our $M_\star$
posterior samples has 0.2\,\Msun\ located at the 5th percentile. While this
does not imply total disagreement between our derived mass and that from the
evolutionary models of \citet{shen25-1}, it suggests that improvements to
either our spectroscopic models (e.g. atomic data) or the evolutionary models
calculated by \citet{shen25-1} are required. Adopting our radius (constrained
independently of the spectroscopic data) and assuming a mass of 0.2\,\Msun\
implies $\log g=5.63$\,dex, which is more than $2\sigma$ below our
spectroscopic measurement. On the other hand, altering the initial temperature,
radius, and chemical profiles assumed by \citet{shen25-1} will lead to
different evolutionary tracks in the $L$-$\Teff$ plane, which may permit higher
masses for \WD. Regardless of whether the masses can be made to agree, an even
more egregious disagreement is seen in terms of age where our \Teff\ and
luminosity imply a 12\,Myr age according to the \citet{shen25-1} tracks, in
vast disagreement with our $4.5_{-0.5}^{+0.4}$\,Myr time of flight.

\citet{wongetal24-1} and \citet{bhatetal25-1} have also performed evolutionary
models for \Dsix\ survivors. The helium white dwarf models of
\citet{wongetal24-1} (see their Figure 8) appear quite different to the CO
models of \citet{shen25-1}, where their tracks decrease in luminosity and
radius over time, with a more gradual reduction in temperature. We note that
for our $\simeq 4.5$\,Myr time of flight, the tracks of \citet{wongetal24-1}
find approximately the correct radius to \WD\ of $\simeq 0.1$\,\Rsun, though at
much lower \Teff. Given that these are helium white dwarf models, it is
unsurprising that they disagree with the apparent CO core composition of \WD.
\citet{bhatetal25-1} investigated the inflated nature of the known \Dsix\
remnants finding that the supernova shock heating could not maintain the
inflated radii of the donors for more than $\sim 10^3$\,yr -- much shorter than
the estimated age of \WD\ and the other known \Dsix\ stars, implying some other
mechanism is responsible for the large radii of \Dsix\ remnants. At the
$\simeq4.5$\,Myr age for \WD\ their models predict a radius that is an order of
magnitude too low and a temperature several times too high (see their Figure
11), though these discrepancies cancel, yielding a luminosity estimate
approximately correct for the age of \WD. In summary, the disagreements between
these evolutionary calculations and the results we present here motivate
further improvements to both the atmospheric and evolutionary models in order
to resolve these tensions.

\subsection{Constraining the ejection velocity from the donor mass}

The velocity of the ejected donor reflects the orbital velocity at the time of
the explosion,\footnotemark\ which is a function of the mass of both components
in the binary. This idea has been explored by \citet{baueretal21-1}, and offers
the potential to be applied to our kinematic analysis of \WD\ for refined
results. Though the parameters we find here are more precise than those in
Fit~1, they are also subject to the caveat of the unknown donor temperature,
and so our results from Fit~1 should still be taken as our most reliable
estimates.

\footnotetext{In principle the $v_\mathrm{ej}$ is a combination of
$v_\mathrm{orb}$ and any kick, $v_\mathrm{kick}$, imparted by the supernova,
and so $v_\mathrm{ej}^2 = v_\mathrm{orb}^2 + v_\mathrm{kick}^2$ (since any
outward kick will be perpendicular to the direction of orbital motion). Unless
the donor mass is particularly low (in which case $v_\mathrm{orb}$ will also be
small), then the orbital component will dominate and the approximation
$|v_\mathrm{ej}| \approx |v_\mathrm{orb}|$ is valid (see Figure~15 of
\citealt{el-badryetal23-1}).}

Combining Kepler's third law, and the Roche lobe radius of the donor
\citep{eggleton83-1}, \citet{baueretal21-1} showed that at the time of
explosion, the orbital speed $v_\mathrm{orb}$ of the donor can be found via
\begin{equation}
    v_\mathrm{orb}^2 = \frac{0.49q^{2/3}GM_\mathrm{acc}}{R_\mathrm{don}(1+q)[0.6q^{2/3}+\ln(1+q^{1/3})]},
\end{equation}
where $G$ is the gravitational constant, $M_\mathrm{acc}$ is the mass of the
accreting primary, $q$ is the mass ratio (i.e.
$q=M_\mathrm{don}/M_\mathrm{acc}$), and $R_\mathrm{don}$ is the donor radius.
Theory indicates that the \Dsix\ detonations should occur for primary masses
between 0.85--1.15\,\Msun, providing lower and upper-limits for
$M_\mathrm{acc}$ (note that the accretor must also be at least as massive as
the donor). Although the donor may have lost some fraction of its mass, it is
still a white dwarf and should therefore still follow a standard white dwarf
mass radius relation up to the detonation. Therefore $R_\mathrm{don}$ is a
function of $M_\mathrm{don}$, and so $v_\mathrm{orb}$ is also a function of
$M_\mathrm{don}$ for both the lower and upper bounds of $M_\mathrm{acc}$. An
important caveat is that $R_\mathrm{don}$ is not just a function of
$M_\mathrm{don}$, but also the effective temperature of the donor, and the
atmospheric composition. In the work of \citet{baueretal21-1}, they
investigated a range of \Teff\ and for both hydrogen and helium dominated
atmospheres. For our analysis we assumed a helium dominated atmosphere as the
dynamical instability leading to \Dsix\ supernovae should not set in until
after any hydrogen envelope is transferred. Similarly, the mass-radius relation
for white dwarfs with He surfaces can still be sensitive to $\Teff$, but
dynamical instability is most likely to set in and generate a supernova for
more compact configurations with $\Teff \lesssim 30{,}000\,\rm K$, and the
mass-radius relation is not particularly sensitive to temperature in this
regime. Therefore, with no other obvious value available, we simply adopted the
current spectroscopic $\Teff$ of 15\,680\,K.

We used the \citet{bedardetal20-1} mass-radius relation to determine the
$R_\mathrm{don}$ from $M_\mathrm{don}$, and the current $\Teff$ of \WD. At this
$\Teff$ the \citet{bedardetal20-1} model grids provide valid results down to
masses of 0.2\,\Msun, which is already well covered by the posterior values of
$M_\mathrm{don}$ in Figure~\ref{fig:corner_traceback}. The upper and lower
bounds on $v_\mathrm{orb}$ as a function of $M_\mathrm{don}$ were implemented
as a prior which is flat between the bounds, but zero elsewhere. Here the
reason to include $f$ as a free parameter becomes apparent as this prior
depends explicitly on $M_\mathrm{don}$, but our likelihood terms for the
surface gravity, and the gravitational redshift component of the spectroscopic
radial velocity depend on the \emph{current} mass of \WD, $M_\star$. Therefore
it is important to include a term representing any potential mass loss arising
from interaction with the supernova ejecta, rather than assume these two masses
to be identical.

\begin{figure}
    \centering
    \includegraphics[width=\columnwidth]{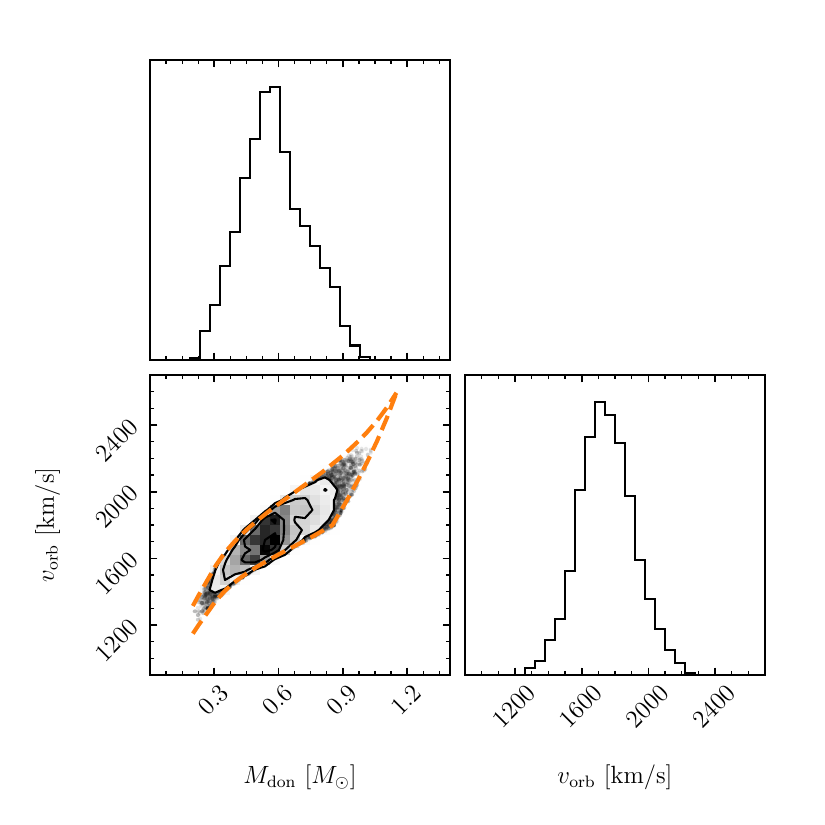}
    \caption{Corner plot for $v_\mathrm{orb}$ and $M_\mathrm{don}$. The bounds
    inferred for $\Teff=15\,680$\,K are shown by the dashed lines.}
    \label{fig:vorb}
\end{figure}

The results from this fit are listed as Fit~2 in Table~\ref{tab:kinematics}.
The joint prior on $v_\mathrm{orb}$ and $M_\mathrm{don}$ limited the parameter
space those quantities can occupy (Figure~\ref{fig:vorb}) improving the
ejection velocity by around a factor of 2 (noting that we make the
approximation $v_\mathrm{ej} = v_\mathrm{orb}$). Consequently several
parameters that are sensitive to the ejection speed, have seen their
uncertainties reduced by a similar amount, in particular $D$, $R_\star$, and
the components of $\mathbf{v}$. The donor mass, $M_\mathrm{don}$ sees a
comparatively marginal improvement, with similar small improvements seen for
$f$ and the derived $M_\star$. The time of flight, $t$, is minimally affected
as it predominantly reflects the scale height of the disc at the inferred
ejection site. The parameters \Teff\ and \ebv\ show no improvement as the
former is constrained only by its prior, and the latter only by the photometry,
which remain unchanged here.

\begin{figure*}
    \centering
    \includegraphics[width=\textwidth]{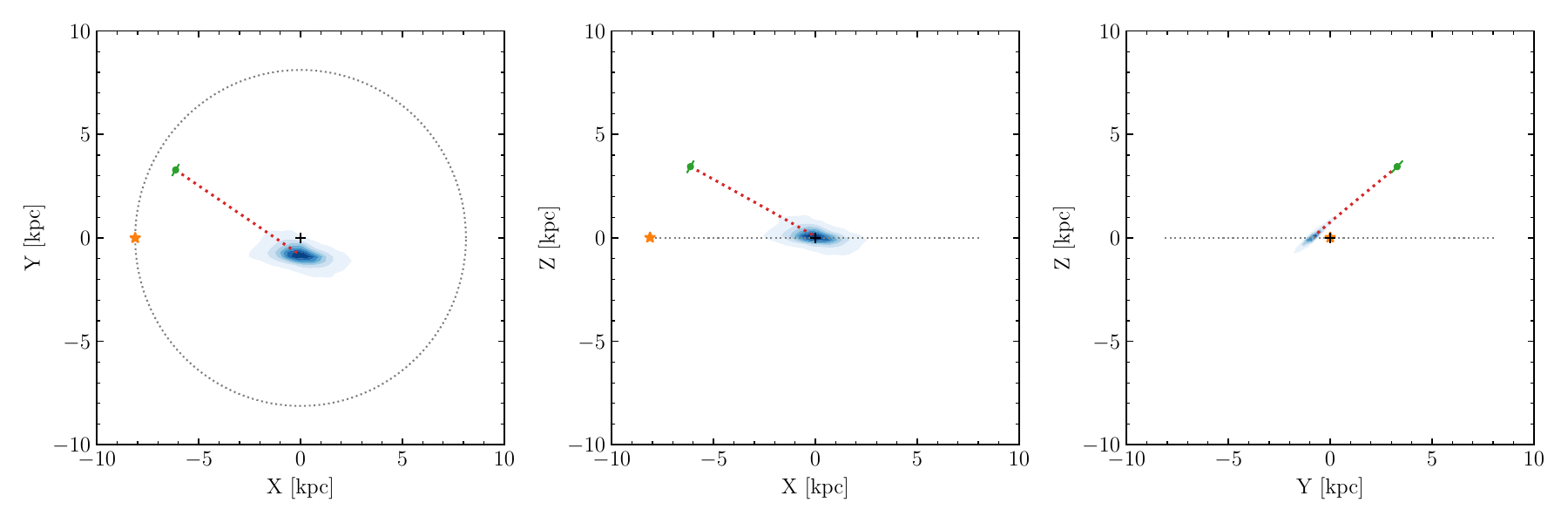}
    \caption{Ejection site and trajectory of \WD\ using posterior parameters
    from Fit 2. All symbols have the same meaning as in
    Figure~\ref{fig:birthsite2}.}
    \label{fig:birthsite2}
\end{figure*}

The inferred location of the ejection site (Figure~\ref{fig:birthsite2}) is
also improved though with most of the improvement in $X_0$, which also leads to
increased correlation between $X_0$ and $Y_0$, as well as $X_0$ and $Z_0$
($\rho_{XY}$ and $\rho_{XZ}$ in Table~\ref{tab:kinematics}). Once more, the
posterior volume covered by these spatial parameters does not overlap the
Galactic centre due to the strong correlation between $Y_0$ and $Z_0$.

Employing these constraints, which are physically motivated by the properties
of the binary at the time of explosion, greatly improves the precision of many
quantities of interest. However, as this analysis depends on an arbitrarily
chosen value of the donor \Teff\ at the explosion epoch, we reiterate that
these results may be subject to additional uncertainties in the donor radius at
time of explosion, and that the results from Fit~1 should be used when citing
this work. Furthermore, \citet{braudo+soker24-1} has shown that the ejection
velocity of the donor is typically reduced by 8--11\,percent compared to its
orbital velocity (due to the finite velocity of the supernova ejecta), an
effect that has not been taken into account here, and may imply an even more
massive value of $M_\mathrm{don}$.

\subsection{Future work for kinematic analyses}
\label{sec:future}

While the analyses performed throughout this section have been allowed us to
determine several key properties of \WD\ (its ejection site,  ejection
velocity, and time of flight), our methodology is not without its limitations.
Firstly, our distance prior, $P(D) \propto D^2$, assumes a constant density of
objects along the line of sight. This is likely to be true close to the
Galactic plane, but will inevitably drop off at distances beyond a few kpc. At
larger distances, the number of \emph{detectable} \Dsix\ stars will decrease as
their apparent brightness will be too low for \emph{Gaia}, but will depend on
the evolutionary state of the specific object. A realistic distance prior will
therefore require simulating the population of \Dsix\ stars along different
sight lines, considering for the distribution of ejection velocities, and
potentially incorporating the evolving luminosity of the ejected donors
\citep{shen25-1}, if the disagreements presented above can be resolved. We
therefore intend to explore this line of research in future work, though note
that a preliminary investigation suggests the density of \Dsix\ stars does not
significantly drop for distances within 10\,kpc along the sight line of \WD.

As a further aside, our model presented earlier in this section integrates the
path of \WD\ back towards a potential origin. However it is also possible to
construct a forward model, where instead of the current distance as a free
parameter, the ejection site $\mathbf{r}_0$ is used instead, and instead of the
current velocity components, the velocity at ejection
($\mathbf{v}_\mathrm{ej}$) is used. Clearly this approach introduces two
additional free-parameters to the model, but also requires that the astrometric
term in the likelihood include the \emph{Gaia} RA and Dec and their
uncertainties, otherwise the orbits are permitted to terminate at any sky
coordinate. Such a model is attractive, as it completely removes the need for a
distance prior, with the prior on $\mathbf{r}_0$ simply being the stellar
density. While such a model is simple to implement, in practice we found that
it was not possible to properly sample the posterior distribution, due to the
covariance between parameters becoming almost singular. Consider the case that
$\mathbf{r}_0$ is fixed: in order for the trajectory of the ejected donor to
terminate within the cone described by the RA, Dec and uncertainties (sub-mas),
extremely fine-tune parameters are needed for $\mathbf{v}_\mathrm{ej}$ and $t$,
occupying a miniscule volume within the parameter space. A moderate change in
$\mathbf{r}_0$ of say 0.5\,kpc, will again lead to extremely fine-tuned, though
substantially different parameters in order to once again land within the
target volume defined by the astrometric likelihood. Therefore, when
$\mathbf{r}_0$ is allowed to be free, the posterior density of $\mathbf{r}_0$,
$\mathbf{v}_\mathrm{ej}$, and $t$ essentially exists along a hyper-surface
within the posterior hyper-volume. In our attempted implementation of the
forward model, we were unable to find a way to deal with this correlation
without affecting the posterior more generally: We tried artificially inflating
the RA and Dec uncertainties, in order to increase the volume defined by the
likelihood, but without substantially affecting the length of the integrated
orbit (compared to the presumed uncertainty in $\mathbf{r}_0$). While this did
reduce the extent of the parameter correlation, it also scaled the posterior
for parameters relating to the orbit, and so no optimal value could be found.
This is not to say that the forward model cannot successfully be implemented,
but goes beyond the scope of this present work when adequate results can
already be obtained from the reverse model.

Finally, the approach we have developed here can be applied more generally to
other bound remnants of supernovae including both the \Dsix\ and \lp\ stars.
The analysis in this work represents an ideal case where \WD\ is high above the
Galactic plane, a spectroscopic $\log g$ has been measured to restrict
$M_\star$ and $R_\star$, and a best fitting spectral model is available to
compare against the photometry and constrain \ebv. For objects without
spectroscopic analyses, as long as a full set of astrometry is available
(including the radial velocity), it is still possible to constrain information
about the ejection site, ejection velocity, and time of flight, though other
important quantities such as a the mass and radius will be inaccessible without
further information. Analyses of other runaway stars is beyond the scope of
this work, but will be investigated in the future.

\section{Conclusions}
\label{sec:conc}

\WDlong\ was previously identified from its peculiar spectrum hinting at a
supernova origin, where we have now established it the surviving donor of a
\Dsix\ type Ia supernova. Using our GTC observations, we performed the most
detailed spectroscopic analysis of any \Dsix\ star to date, measuring the
abundances for eight different elements. This revealed a C+O dominated
atmosphere rich in IMEs such as Si, S, and Ca -- a composition found to be
consistent with an exposed CO white dwarf core enhanced by a few percent in the
IME-rich ejecta of a double-detonation supernova. Despite a low-precision
parallax, we were still able to infer a distance to \WD\ of $\approx 5$\,kpc
and a Galactocentric velocity of $\approx 1900$\,\kms, corroborating the \Dsix\
explanation for \WD. We then extended our analysis to explore the origin and
kinematics of \WD in detail by tracing back its orbit, refining the distance
and velocity measurements, and finding a time of flight of
$4.5_{-0.5}^{+0.4}$\,Myr. We found that \WD\ originated from the inner few kpc
of the Galaxy, though the ejection site does not overlap the Galactic centre,
and favours ejection from the disc population as opposed to the bulge. Finally,
we refined this model further by restricting the ejection-speed based on the
mass of the donor, as suggested by theoretical work, which we found led to even
tighter constraints on the ejection site, ejection velocity and current
distance from the Sun.

\section*{Acknowledgements}

Based on observations made with the Gran Telescopio Canarias (GTC), installed
in the Spanish Observatorio del Roque de los Muchachos of the Instituto de
Astrof´ısica de Canarias (IAC), in the island of La Palma.     This work is
(partly) based on data obtained with the instrument OSIRIS, built by a
Consortium led by the Instituto de Astrof\'sica de Canarias in collaboration
with the Instituto de Astronomía of the Universidad Autónoma de México. OSIRIS
was funded by GRANTECAN and the National Plan of Astronomy and Astrophysics of
the Spanish Government. The Starlink software \citep{currieetal14-1} is
currently supported by the East Asian Observatory. This work made use of
Astropy: (\url{https://www.astropy.org}) a community-developed core Python
package and an ecosystem of tools and resources for astronomy
\citep{astropy13-1,astropy18-1,astropy22-1}. MAH acknowledges a Warwick
Astrophysics prize post-doctoral fellowship made possible thanks to a generous
philanthropic donation. Support for KJS was provided by NASA through the
Astrophysics Theory Program (80NSSC20K0544) and by NASA/ESA Hubble Space
Telescope program Nos.\ 15871, 15918, and 17441. RR acknowledges support from
Grant RYC2021-030837-I, funded by MCIN/AEI/ 10.13039/501100011033 and by
``European Union NextGeneration EU/PRTR''. This project has received funding
from the European Research Council (ERC) under the European Union’s Horizon
2020 research and innovation programme (Grant agreement No. 101020057). This
work was partially supported by the Spanish MINECO grant PID2023-148661NB-I00
and by the AGAUR/Generalitat de Catalunya grant SGR-386/2021. MAH acknowledges
useful discussions with Stuart Littlefair on statistical methods, and Detlev
Koester regarding their model atmosphere code. We thank the anonymous referee
for their constructive feedback which improved the quality of this work.

%%%%%%%%%%%%%%%%%%%%%%%%%%%%%%%%%%%%%%%%%%%%%%%%%%
\section*{Data Availability}

All photometry and astrometry used in this work are publicly available
(\emph{Gaia}, SDSS, PanSTARRS). The raw GTC spectra can be found on the GTC
archive with programme number GTC71-19A.



\bibliographystyle{mnras}
\bibliography{aamnem99,aabib} % if your bibtex file is called example.bib

\begin{thebibliography}{}
\makeatletter
\relax
\def\mn@urlcharsother{\let\do\@makeother \do\$\do\&\do\#\do\^\do\_\do\%\do\~}
\def\mn@doi{\begingroup\mn@urlcharsother \@ifnextchar [ {\mn@doi@}
  {\mn@doi@[]}}
\def\mn@doi@[#1]#2{\def\@tempa{#1}\ifx\@tempa\@empty \href
  {http://dx.doi.org/#2} {doi:#2}\else \href {http://dx.doi.org/#2} {#1}\fi
  \endgroup}
\def\mn@eprint#1#2{\mn@eprint@#1:#2::\@nil}
\def\mn@eprint@arXiv#1{\href {http://arxiv.org/abs/#1} {{\tt arXiv:#1}}}
\def\mn@eprint@dblp#1{\href {http://dblp.uni-trier.de/rec/bibtex/#1.xml}
  {dblp:#1}}
\def\mn@eprint@#1:#2:#3:#4\@nil{\def\@tempa {#1}\def\@tempb {#2}\def\@tempc
  {#3}\ifx \@tempc \@empty \let \@tempc \@tempb \let \@tempb \@tempa \fi \ifx
  \@tempb \@empty \def\@tempb {arXiv}\fi \@ifundefined
  {mn@eprint@\@tempb}{\@tempb:\@tempc}{\expandafter \expandafter \csname
  mn@eprint@\@tempb\endcsname \expandafter{\@tempc}}}

\bibitem[\protect\citeauthoryear{{Astropy Collaboration} et~al.,}{{Astropy
  Collaboration} et~al.}{2013}]{astropy13-1}
{Astropy Collaboration} et~al., 2013, \mn@doi [A\&A]
  {10.1051/0004-6361/201322068}, \href
  {https://ui.adsabs.harvard.edu/abs/2013A&A...558A..33A} {558, A33}

\bibitem[\protect\citeauthoryear{{Astropy Collaboration} et~al.,}{{Astropy
  Collaboration} et~al.}{2018}]{astropy18-1}
{Astropy Collaboration} et~al., 2018, \mn@doi [AJ] {10.3847/1538-3881/aabc4f},
  \href {https://ui.adsabs.harvard.edu/abs/2018AJ....156..123A} {156, 123}

\bibitem[\protect\citeauthoryear{{Astropy Collaboration} et~al.,}{{Astropy
  Collaboration} et~al.}{2022}]{astropy22-1}
{Astropy Collaboration} et~al., 2022, \mn@doi [ApJ Lett.]
  {10.3847/1538-4357/ac7c74}, \href
  {https://ui.adsabs.harvard.edu/abs/2022ApJ...935..167A} {935, 167}

\bibitem[\protect\citeauthoryear{{Bauer}, {Chandra}, {Shen}  \&
  {Hermes}}{{Bauer} et~al.}{2021}]{baueretal21-1}
{Bauer} E.~B.,  {Chandra} V.,  {Shen} K.~J.,   {Hermes} J.~J.,  2021, \mn@doi
  [ApJ Lett.] {10.3847/2041-8213/ac432d}, \href
  {https://ui.adsabs.harvard.edu/abs/2021ApJ...923L..34B} {923, L34}

\bibitem[\protect\citeauthoryear{{B{\'e}dard}, {Bergeron}, {Brassard}  \&
  {Fontaine}}{{B{\'e}dard} et~al.}{2020}]{bedardetal20-1}
{B{\'e}dard} A.,  {Bergeron} P.,  {Brassard} P.,   {Fontaine} G.,  2020,
  \mn@doi [ApJ] {10.3847/1538-4357/abafbe}, \href
  {https://ui.adsabs.harvard.edu/abs/2020ApJ...901...93B} {901, 93}

\bibitem[\protect\citeauthoryear{{Bell}, {Berry}, {Graves}, {Currie}  \&
  {Draper}}{{Bell} et~al.}{2024}]{belletal24-1}
{Bell} G.~S.,  {Berry} D.~S.,  {Graves} S.~F.,  {Currie} M.~J.,   {Draper}
  P.~W.,  2024, in {Hugo} B.~V.,  {Van Rooyen} R.,   {Smirnov} O.~M.,  eds,
  Astronomical Society of the Pacific Conference Series Vol. 535, Astromical
  Data Analysis Software and Systems XXXI. p.~455

\bibitem[\protect\citeauthoryear{{Bhat}, {Bauer}, {Pakmor}, {Shen}, {Caiazzo},
  {Rajamuthukumar}, {El-Badry}  \& {Kerzendorf}}{{Bhat}
  et~al.}{2025}]{bhatetal25-1}
{Bhat} A.,  {Bauer} E.~B.,  {Pakmor} R.,  {Shen} K.~J.,  {Caiazzo} I.,
  {Rajamuthukumar} A.~S.,  {El-Badry} K.,   {Kerzendorf} W.~E.,  2025, \mn@doi
  [A\&A] {10.1051/0004-6361/202451371}, \href
  {https://ui.adsabs.harvard.edu/abs/2025A&A...693A.114B} {693, A114}

\bibitem[\protect\citeauthoryear{{Boos}, {Townsley}, {Shen}, {Caldwell}  \&
  {Miles}}{{Boos} et~al.}{2021}]{boosetal21-1}
{Boos} S.~J.,  {Townsley} D.~M.,  {Shen} K.~J.,  {Caldwell} S.,   {Miles}
  B.~J.,  2021, \mn@doi [ApJ] {10.3847/1538-4357/ac07a2}, \href
  {https://ui.adsabs.harvard.edu/abs/2021ApJ...919..126B} {919, 126}

\bibitem[\protect\citeauthoryear{{Boos}, {Townsley}  \& {Shen}}{{Boos}
  et~al.}{2024}]{boosetal24-1}
{Boos} S.~J.,  {Townsley} D.~M.,   {Shen} K.~J.,  2024, \mn@doi [ApJ]
  {10.3847/1538-4357/ad5da2}, \href
  {https://ui.adsabs.harvard.edu/abs/2024ApJ...972..200B} {972, 200}

\bibitem[\protect\citeauthoryear{{Bovy}}{{Bovy}}{2015}]{bovyetal15-1}
{Bovy} J.,  2015, \mn@doi [ApJS] {10.1088/0067-0049/216/2/29}, \href
  {https://ui.adsabs.harvard.edu/abs/2015ApJS..216...29B} {216, 29}

\bibitem[\protect\citeauthoryear{{Braudo} \& {Soker}}{{Braudo} \&
  {Soker}}{2024}]{braudo+soker24-1}
{Braudo} J.,  {Soker} N.,  2024, \mn@doi [The Open Journal of Astrophysics]
  {10.21105/astro.2310.16554}, \href
  {https://ui.adsabs.harvard.edu/abs/2024OJAp....7E...7B} {7, 7}

\bibitem[\protect\citeauthoryear{{Chandra} et~al.,}{{Chandra}
  et~al.}{2022}]{chandraetal22-1}
{Chandra} V.,  et~al., 2022, \mn@doi [MNRAS] {10.1093/mnras/stac883}, \href
  {https://ui.adsabs.harvard.edu/abs/2022MNRAS.512.6122C} {512, 6122}

\bibitem[\protect\citeauthoryear{{Cunto}, {Mendoza}, {Ochsenbein}  \&
  {Zeippen}}{{Cunto} et~al.}{1993}]{cuntoetal93-1}
{Cunto} W.,  {Mendoza} C.,  {Ochsenbein} F.,   {Zeippen} C.~J.,  1993, A\&A,
  \href {1993A&A...275L...5C} {275, L5}

\bibitem[\protect\citeauthoryear{{Currie}, {Berry}, {Jenness}, {Gibb}, {Bell}
  \& {Draper}}{{Currie} et~al.}{2014}]{currieetal14-1}
{Currie} M.~J.,  {Berry} D.~S.,  {Jenness} T.,  {Gibb} A.~G.,  {Bell} G.~S.,
  {Draper} P.~W.,  2014, in {Manset} N.,  {Forshay} P.,  eds,  Astronomical
  Society of the Pacific Conference Series Vol. 485, Astronomical Data Analysis
  Software and Systems XXIII. p.~391

\bibitem[\protect\citeauthoryear{{Dan}, {Rosswog}, {Guillochon}  \&
  {Ramirez-Ruiz}}{{Dan} et~al.}{2011}]{danetal11-1}
{Dan} M.,  {Rosswog} S.,  {Guillochon} J.,   {Ramirez-Ruiz} E.,  2011, \mn@doi
  [ApJ] {10.1088/0004-637X/737/2/89}, \href
  {https://ui.adsabs.harvard.edu/abs/2011ApJ...737...89D} {737, 89}

\bibitem[\protect\citeauthoryear{{Eggleton}}{{Eggleton}}{1983}]{eggleton83-1}
{Eggleton} P.~P.,  1983, ApJ, \href {1983ApJ...268..368E} {268, 368}

\bibitem[\protect\citeauthoryear{{El-Badry} et~al.,}{{El-Badry}
  et~al.}{2023}]{el-badryetal23-1}
{El-Badry} K.,  et~al., 2023, \mn@doi [The Open Journal of Astrophysics]
  {10.21105/astro.2306.03914}, \href
  {https://ui.adsabs.harvard.edu/abs/2023OJAp....6E..28E} {6, 28}

\bibitem[\protect\citeauthoryear{{Evans}, {Renzo}  \& {Rossi}}{{Evans}
  et~al.}{2020}]{evansetal20-1}
{Evans} F.~A.,  {Renzo} M.,   {Rossi} E.~M.,  2020, \mn@doi [MNRAS]
  {10.1093/mnras/staa2334}, \href
  {https://ui.adsabs.harvard.edu/abs/2020MNRAS.497.5344E} {497, 5344}

\bibitem[\protect\citeauthoryear{{Fink}, {Hillebrandt}  \& {R{\"o}pke}}{{Fink}
  et~al.}{2007}]{finketal07-1}
{Fink} M.,  {Hillebrandt} W.,   {R{\"o}pke} F.~K.,  2007, \mn@doi [A\&A]
  {10.1051/0004-6361:20078438}, \href
  {https://ui.adsabs.harvard.edu/abs/2007A&A...476.1133F} {476, 1133}

\bibitem[\protect\citeauthoryear{{Fink}, {R{\"o}pke}, {Hillebrandt},
  {Seitenzahl}, {Sim}  \& {Kromer}}{{Fink} et~al.}{2010}]{finketal10-1}
{Fink} M.,  {R{\"o}pke} F.~K.,  {Hillebrandt} W.,  {Seitenzahl} I.~R.,  {Sim}
  S.~A.,   {Kromer} M.,  2010, \mn@doi [A\&A] {10.1051/0004-6361/200913892},
  \href {https://ui.adsabs.harvard.edu/abs/2010A&A...514A..53F} {514, A53}

\bibitem[\protect\citeauthoryear{{Fink} et~al.,}{{Fink}
  et~al.}{2014}]{finketal14-1}
{Fink} M.,  et~al., 2014, \mn@doi [MNRAS] {10.1093/mnras/stt2315}, \href
  {https://ui.adsabs.harvard.edu/abs/2014MNRAS.438.1762F} {438, 1762}

\bibitem[\protect\citeauthoryear{{Foley} et~al.,}{{Foley}
  et~al.}{2013}]{foleyetal13-1}
{Foley} R.~J.,  et~al., 2013, \mn@doi [ApJ] {10.1088/0004-637X/767/1/57}, \href
  {https://ui.adsabs.harvard.edu/abs/2013ApJ...767...57F} {767, 57}

\bibitem[\protect\citeauthoryear{{Foreman-Mackey}, {Hogg}, {Lang}  \&
  {Goodman}}{{Foreman-Mackey} et~al.}{2013}]{foremanmackeyetal13-1}
{Foreman-Mackey} D.,  {Hogg} D.~W.,  {Lang} D.,   {Goodman} J.,  2013, \mn@doi
  [PASP] {10.1086/670067}, \href
  {http://adsabs.harvard.edu/abs/2013PASP..125..306F} {125, 306}

\bibitem[\protect\citeauthoryear{{GRAVITY Collaboration} et~al.,}{{GRAVITY
  Collaboration} et~al.}{2018}]{gravitycollab18-1}
{GRAVITY Collaboration} et~al., 2018, \mn@doi [A\&A]
  {10.1051/0004-6361/201833718}, \href
  {https://ui.adsabs.harvard.edu/abs/2018A&A...615L..15G} {615, L15}

\bibitem[\protect\citeauthoryear{{G{\"a}nsicke}, {Koester}, {Raddi}, {Toloza}
  \& {Kepler}}{{G{\"a}nsicke} et~al.}{2020}]{gaensickeetal20-1}
{G{\"a}nsicke} B.~T.,  {Koester} D.,  {Raddi} R.,  {Toloza} O.,   {Kepler}
  S.~O.,  2020, \mn@doi [MNRAS] {10.1093/mnras/staa1761}, \href
  {https://ui.adsabs.harvard.edu/abs/2020MNRAS.496.4079G} {496, 4079}

\bibitem[\protect\citeauthoryear{{Geier} et~al.,}{{Geier}
  et~al.}{2015}]{geieretal15-1}
{Geier} S.,  et~al., 2015, \mn@doi [Sci] {10.1126/science.1259063}, \href
  {http://adsabs.harvard.edu/abs/2015Sci...347.1126G} {347, 1126}

\bibitem[\protect\citeauthoryear{{Gordon}, {Clayton}, {Decleir}, {Fitzpatrick},
  {Massa}, {Misselt}  \& {Tollerud}}{{Gordon} et~al.}{2023}]{gordonetal23-1}
{Gordon} K.~D.,  {Clayton} G.~C.,  {Decleir} M.,  {Fitzpatrick} E.~L.,  {Massa}
  D.,  {Misselt} K.~A.,   {Tollerud} E.~J.,  2023, \mn@doi [ApJ]
  {10.3847/1538-4357/accb59}, \href
  {https://ui.adsabs.harvard.edu/abs/2023ApJ...950...86G} {950, 86}

\bibitem[\protect\citeauthoryear{{Guillochon}, {Dan}, {Ramirez-Ruiz}  \&
  {Rosswog}}{{Guillochon} et~al.}{2010}]{guillochonetal10-1}
{Guillochon} J.,  {Dan} M.,  {Ramirez-Ruiz} E.,   {Rosswog} S.,  2010, \mn@doi
  [ApJ Lett.] {10.1088/2041-8205/709/1/L64}, \href
  {https://ui.adsabs.harvard.edu/abs/2010ApJ...709L..64G} {709, L64}

\bibitem[\protect\citeauthoryear{{Hermes}, {Putterman}, {Hollands}, {Wilson},
  {Swan}, {Raddi}, {Shen}  \& {G{\"a}nsicke}}{{Hermes}
  et~al.}{2021}]{hermesetal21-1}
{Hermes} J.~J.,  {Putterman} O.,  {Hollands} M.~A.,  {Wilson} D.~J.,  {Swan}
  A.,  {Raddi} R.,  {Shen} K.~J.,   {G{\"a}nsicke} B.~T.,  2021, \mn@doi [ApJ
  Lett.] {10.3847/2041-8213/ac00a8}, \href
  {https://ui.adsabs.harvard.edu/abs/2021ApJ...914L...3H} {914, L3}

\bibitem[\protect\citeauthoryear{{Hills}}{{Hills}}{1988}]{hills88-1}
{Hills} J.~G.,  1988, \mn@doi [Nat] {10.1038/331687a0}, \href
  {https://ui.adsabs.harvard.edu/abs/1988Natur.331..687H} {331, 687}

\bibitem[\protect\citeauthoryear{{Hollands} et~al.,}{{Hollands}
  et~al.}{2020}]{hollandsetal20-1}
{Hollands} M.~A.,  et~al., 2020, \mn@doi [Nat Astron.]
  {10.1038/s41550-020-1028-0}, \href
  {https://ui.adsabs.harvard.edu/abs/2020NatAs...4..663H} {4, 663}

\bibitem[\protect\citeauthoryear{{Horne}}{{Horne}}{1986}]{horne86-1}
{Horne} K.,  1986, PASP, \href {1986PASP...98..609H} {98, 609}

\bibitem[\protect\citeauthoryear{{Igoshev}, {Perets}  \& {Hallakoun}}{{Igoshev}
  et~al.}{2023}]{igoshevetal23-1}
{Igoshev} A.~P.,  {Perets} H.,   {Hallakoun} N.,  2023, \mn@doi [MNRAS]
  {10.1093/mnras/stac3488}, \href
  {https://ui.adsabs.harvard.edu/abs/2023MNRAS.518.6223I} {518, 6223}

\bibitem[\protect\citeauthoryear{{Irrgang}, {Kreuzer}  \& {Heber}}{{Irrgang}
  et~al.}{2018}]{irrgangetal18-1}
{Irrgang} A.,  {Kreuzer} S.,   {Heber} U.,  2018, \mn@doi [A\&A]
  {10.1051/0004-6361/201833874}, \href
  {https://ui.adsabs.harvard.edu/abs/2018A&A...620A..48I} {620, A48}

\bibitem[\protect\citeauthoryear{{Jha}, {Branch}, {Chornock}, {Foley}, {Li},
  {Swift}, {Casebeer}  \& {Filippenko}}{{Jha} et~al.}{2006}]{jhaetal06-1}
{Jha} S.,  {Branch} D.,  {Chornock} R.,  {Foley} R.~J.,  {Li} W.,  {Swift}
  B.~J.,  {Casebeer} D.,   {Filippenko} A.~V.,  2006, \mn@doi [AJ]
  {10.1086/504599}, \href
  {https://ui.adsabs.harvard.edu/abs/2006AJ....132..189J} {132, 189}

\bibitem[\protect\citeauthoryear{{Jones} et~al.,}{{Jones}
  et~al.}{2019}]{jonesetal19-1}
{Jones} S.,  et~al., 2019, \mn@doi [A\&A] {10.1051/0004-6361/201834381}, \href
  {https://ui.adsabs.harvard.edu/abs/2019A&A...622A..74J} {622, A74}

\bibitem[\protect\citeauthoryear{{Karamanis}, {Nabergoj}, {Beutler}, {Peacock}
  \& {Seljak}}{{Karamanis} et~al.}{2022a}]{karamanisetal22-2}
{Karamanis} M.,  {Nabergoj} D.,  {Beutler} F.,  {Peacock} J.,   {Seljak} U.,
  2022a, \mn@doi [J. Open Source Softw.] {10.21105/joss.04634}, \href
  {https://ui.adsabs.harvard.edu/abs/2022JOSS....7.4634K} {7, 4634}

\bibitem[\protect\citeauthoryear{{Karamanis}, {Beutler}, {Peacock}, {Nabergoj}
  \& {Seljak}}{{Karamanis} et~al.}{2022b}]{karamanisetal22-1}
{Karamanis} M.,  {Beutler} F.,  {Peacock} J.~A.,  {Nabergoj} D.,   {Seljak} U.,
   2022b, \mn@doi [MNRAS] {10.1093/mnras/stac2272}, \href
  {https://ui.adsabs.harvard.edu/abs/2022MNRAS.516.1644K} {516, 1644}

\bibitem[\protect\citeauthoryear{{Kepler} et~al.,}{{Kepler}
  et~al.}{2016}]{kepleretal16-1}
{Kepler} S.~O.,  et~al., 2016, \mn@doi [MNRAS] {10.1093/mnras/stv2526}, \href
  {http://adsabs.harvard.edu/abs/2016MNRAS.455.3413K} {455, 3413}

\bibitem[\protect\citeauthoryear{{Khokhlov}}{{Khokhlov}}{1991}]{khokhlov91-1}
{Khokhlov} A.~M.,  1991, A\&A, \href
  {https://ui.adsabs.harvard.edu/abs/1991A&A...245..114K} {245, 114}

\bibitem[\protect\citeauthoryear{{Kilic}, {Bergeron}, {Blouin}, {Jewett},
  {Brown}  \& {Moss}}{{Kilic} et~al.}{2024}]{kilicetal24-1}
{Kilic} M.,  {Bergeron} P.,  {Blouin} S.,  {Jewett} G.,  {Brown} W.~R.,
  {Moss} A.,  2024, \mn@doi [ApJ] {10.3847/1538-4357/ad3440}, \href
  {https://ui.adsabs.harvard.edu/abs/2024ApJ...965..159K} {965, 159}

\bibitem[\protect\citeauthoryear{{Koester}}{{Koester}}{2010}]{koester10-1}
{Koester} D.,  2010, Memorie della Societa Astronomica Italiana,, \href
  {2010MmSAI..81..921K} {81, 921}

\bibitem[\protect\citeauthoryear{{Koposov} et~al.,}{{Koposov}
  et~al.}{2020}]{koposovetal20-1}
{Koposov} S.~E.,  et~al., 2020, \mn@doi [MNRAS] {10.1093/mnras/stz3081}, \href
  {https://ui.adsabs.harvard.edu/abs/2020MNRAS.491.2465K} {491, 2465}

\bibitem[\protect\citeauthoryear{Kramida, {Yu.~Ralchenko}, Reader  \& {and NIST
  ASD Team}}{Kramida et~al.}{2024}]{kramidaetal24-1}
Kramida A.,  {Yu.~Ralchenko} Reader J.,   {and NIST ASD Team} 2024, {NIST
  Atomic Spectra Database (ver. 5.12), [Online]. Available:
  {\tt{https://physics.nist.gov/asd}} [2025, March 13]. National Institute of
  Standards and Technology, Gaithersburg, MD.}

\bibitem[\protect\citeauthoryear{{Kromer} et~al.,}{{Kromer}
  et~al.}{2013}]{kromeretal13-1}
{Kromer} M.,  et~al., 2013, \mn@doi [MNRAS] {10.1093/mnras/sts498}, \href
  {https://ui.adsabs.harvard.edu/abs/2013MNRAS.429.2287K} {429, 2287}

\bibitem[\protect\citeauthoryear{{Li} et~al.,}{{Li} et~al.}{2003}]{lietal03-1}
{Li} W.,  et~al., 2003, \mn@doi [PASP] {10.1086/374200}, \href
  {https://ui.adsabs.harvard.edu/abs/2003PASP..115..453L} {115, 453}

\bibitem[\protect\citeauthoryear{{Livne}}{{Livne}}{1990}]{livne90-1}
{Livne} E.,  1990, \mn@doi [ApJ Lett.] {10.1086/185721}, \href
  {https://ui.adsabs.harvard.edu/abs/1990ApJ...354L..53L} {354, L53}

\bibitem[\protect\citeauthoryear{{Marsh}}{{Marsh}}{1989}]{marsh89-1}
{Marsh} T.~R.,  1989, PASP, \href {1989PASP..101.1032M} {101, 1032}

\bibitem[\protect\citeauthoryear{{McCully} et~al.,}{{McCully}
  et~al.}{2014}]{mccullyetal14-1}
{McCully} C.,  et~al., 2014, \mn@doi [Nat] {10.1038/nature13615}, \href
  {https://ui.adsabs.harvard.edu/abs/2014Natur.512...54M} {512, 54}

\bibitem[\protect\citeauthoryear{{Miyamoto} \& {Nagai}}{{Miyamoto} \&
  {Nagai}}{1975}]{miyamoto+nagai75-1}
{Miyamoto} M.,  {Nagai} R.,  1975, PASJ, \href
  {https://ui.adsabs.harvard.edu/abs/1975PASJ...27..533M} {27, 533}

\bibitem[\protect\citeauthoryear{{Navarro}, {Frenk}  \& {White}}{{Navarro}
  et~al.}{1997}]{navarroetal97-1}
{Navarro} J.~F.,  {Frenk} C.~S.,   {White} S. D.~M.,  1997, \mn@doi [ApJ]
  {10.1086/304888}, \href
  {https://ui.adsabs.harvard.edu/abs/1997ApJ...490..493N} {490, 493}

\bibitem[\protect\citeauthoryear{{O'Brien} et~al.,}{{O'Brien}
  et~al.}{2024}]{obrienetal24-1}
{O'Brien} M.~W.,  et~al., 2024, \mn@doi [MNRAS] {10.1093/mnras/stad3773}, \href
  {https://ui.adsabs.harvard.edu/abs/2024MNRAS.527.8687O} {527, 8687}

\bibitem[\protect\citeauthoryear{{Pakmor}, {Kromer}, {Taubenberger}  \&
  {Springel}}{{Pakmor} et~al.}{2013}]{pakmoretal13-1}
{Pakmor} R.,  {Kromer} M.,  {Taubenberger} S.,   {Springel} V.,  2013, \mn@doi
  [ApJ Lett.] {10.1088/2041-8205/770/1/L8}, \href
  {https://ui.adsabs.harvard.edu/abs/2013ApJ...770L...8P} {770, L8}

\bibitem[\protect\citeauthoryear{{Pakmor} et~al.,}{{Pakmor}
  et~al.}{2022}]{pakmoretal22-1}
{Pakmor} R.,  et~al., 2022, \mn@doi [MNRAS] {10.1093/mnras/stac3107}, \href
  {https://ui.adsabs.harvard.edu/abs/2022MNRAS.517.5260P} {517, 5260}

\bibitem[\protect\citeauthoryear{{Perlmutter} et~al.,}{{Perlmutter}
  et~al.}{1999}]{perlmutteretal99-1}
{Perlmutter} S.,  et~al., 1999, \mn@doi [ApJ] {10.1086/307221}, \href
  {1999ApJ...517..565P} {517, 565}

\bibitem[\protect\citeauthoryear{{Raddi}, {Hollands}, {Koester},
  {G{\"a}nsicke}, {Gentile Fusillo}, {Hermes}  \& {Townsley}}{{Raddi}
  et~al.}{2018a}]{raddietal18-1}
{Raddi} R.,  {Hollands} M.~A.,  {Koester} D.,  {G{\"a}nsicke} B.~T.,  {Gentile
  Fusillo} N.~P.,  {Hermes} J.~J.,   {Townsley} D.~M.,  2018a, \mn@doi [ApJ]
  {10.3847/1538-4357/aab899}, \href
  {https://ui.adsabs.harvard.edu/abs/2018ApJ...858....3R} {858, 3}

\bibitem[\protect\citeauthoryear{{Raddi}, {Hollands}, {G{\"a}nsicke},
  {Townsley}, {Hermes}, {Gentile Fusillo}  \& {Koester}}{{Raddi}
  et~al.}{2018b}]{raddietal18-2}
{Raddi} R.,  {Hollands} M.~A.,  {G{\"a}nsicke} B.~T.,  {Townsley} D.~M.,
  {Hermes} J.~J.,  {Gentile Fusillo} N.~P.,   {Koester} D.,  2018b, \mn@doi
  [MNRAS] {10.1093/mnrasl/sly103}, \href
  {https://ui.adsabs.harvard.edu/abs/2018MNRAS.479L..96R} {\hspace{0pt}479,
  L96}

\bibitem[\protect\citeauthoryear{{Raddi} et~al.,}{{Raddi}
  et~al.}{2019}]{raddietal19-1}
{Raddi} R.,  et~al., 2019, \mn@doi [MNRAS] {10.1093/mnras/stz1618}, \href
  {https://ui.adsabs.harvard.edu/abs/2019MNRAS.489.1489R} {489, 1489}

\bibitem[\protect\citeauthoryear{{Raskin}, {Scannapieco}, {Fryer},
  {Rockefeller}  \& {Timmes}}{{Raskin} et~al.}{2012}]{raskinetal12-1}
{Raskin} C.,  {Scannapieco} E.,  {Fryer} C.,  {Rockefeller} G.,   {Timmes}
  F.~X.,  2012, \mn@doi [ApJ] {10.1088/0004-637X/746/1/62}, \href
  {https://ui.adsabs.harvard.edu/abs/2012ApJ...746...62R} {746, 62}

\bibitem[\protect\citeauthoryear{{Riess} et~al.,}{{Riess}
  et~al.}{1998}]{riessetal98-1}
{Riess} A.~G.,  et~al., 1998, \mn@doi [AJ] {10.1086/300499}, \href
  {1998AJ....116.1009R} {116, 1009}

\bibitem[\protect\citeauthoryear{{Ruffini} \& {Casey}}{{Ruffini} \&
  {Casey}}{2019}]{ruffinietal19-1}
{Ruffini} N.~J.,  {Casey} A.~R.,  2019, \mn@doi [MNRAS]
  {10.1093/mnras/stz2176}, \href
  {https://ui.adsabs.harvard.edu/abs/2019MNRAS.489..420R} {489, 420}

\bibitem[\protect\citeauthoryear{{Ruiter} \& {Seitenzahl}}{{Ruiter} \&
  {Seitenzahl}}{2025}]{ruiter+seitenzahl25-1}
{Ruiter} A.~J.,  {Seitenzahl} I.~R.,  2025, \mn@doi [A\&AR]
  {10.1007/s00159-024-00158-9}, \href
  {https://ui.adsabs.harvard.edu/abs/2025A&ARv..33....1R} {33, 1}

\bibitem[\protect\citeauthoryear{{Scholz}}{{Scholz}}{2024}]{scholz24-1}
{Scholz} R.~D.,  2024, \mn@doi [A\&A] {10.1051/0004-6361/202348430}, \href
  {https://ui.adsabs.harvard.edu/abs/2024A&A...685A.162S} {685, A162}

\bibitem[\protect\citeauthoryear{{Seitenzahl}, {Cescutti}, {R{\"o}pke},
  {Ruiter}  \& {Pakmor}}{{Seitenzahl} et~al.}{2013}]{seitenzahletal13-2}
{Seitenzahl} I.~R.,  {Cescutti} G.,  {R{\"o}pke} F.~K.,  {Ruiter} A.~J.,
  {Pakmor} R.,  2013, \mn@doi [A\&A] {10.1051/0004-6361/201322599}, \href
  {https://ui.adsabs.harvard.edu/abs/2013A&A...559L...5S} {559, L5}

\bibitem[\protect\citeauthoryear{{Shen}}{{Shen}}{2025}]{shen25-1}
{Shen} K.~J.,  2025, \mn@doi [ApJ] {10.3847/1538-4357/adb42e}, \href
  {https://ui.adsabs.harvard.edu/abs/2025ApJ...982....6S} {982, 6}

\bibitem[\protect\citeauthoryear{{Shen} \& {Bildsten}}{{Shen} \&
  {Bildsten}}{2014}]{shen+bildsten14-1}
{Shen} K.~J.,  {Bildsten} L.,  2014, \mn@doi [ApJ]
  {10.1088/0004-637X/785/1/61}, \href
  {https://ui.adsabs.harvard.edu/abs/2014ApJ...785...61S} {785, 61}

\bibitem[\protect\citeauthoryear{{Shen} \& {Schwab}}{{Shen} \&
  {Schwab}}{2017}]{shen+schwab17-1}
{Shen} K.~J.,  {Schwab} J.,  2017, \mn@doi [ApJ] {10.3847/1538-4357/834/2/180},
  \href {https://ui.adsabs.harvard.edu/abs/2017ApJ...834..180S} {834, 180}

\bibitem[\protect\citeauthoryear{{Shen}, {Idan}  \& {Bildsten}}{{Shen}
  et~al.}{2009}]{shenetal09-1}
{Shen} K.~J.,  {Idan} I.,   {Bildsten} L.,  2009, \mn@doi [ApJ]
  {10.1088/0004-637X/705/1/693}, \href {2009ApJ...705..693S} {705, 693}

\bibitem[\protect\citeauthoryear{{Shen} et~al.,}{{Shen}
  et~al.}{2018}]{shenetal18-1}
{Shen} K.~J.,  et~al., 2018, \mn@doi [ApJ] {10.3847/1538-4357/aad55b}, \href
  {https://ui.adsabs.harvard.edu/abs/2018ApJ...865...15S} {865, 15}

\bibitem[\protect\citeauthoryear{{Shen}, {Boos}  \& {Townsley}}{{Shen}
  et~al.}{2024}]{shenetal24-1}
{Shen} K.~J.,  {Boos} S.~J.,   {Townsley} D.~M.,  2024, \mn@doi [ApJ]
  {10.3847/1538-4357/ad7379}, \href
  {https://ui.adsabs.harvard.edu/abs/2024ApJ...975..127S} {975, 127}

\bibitem[\protect\citeauthoryear{{Shields}, {Arunachalam}, {Kerzendorf},
  {Hughes}, {Biriouk}, {Monk}  \& {Buchner}}{{Shields}
  et~al.}{2023}]{shieldsetal23-1}
{Shields} J.~V.,  {Arunachalam} P.,  {Kerzendorf} W.,  {Hughes} J.~P.,
  {Biriouk} S.,  {Monk} H.,   {Buchner} J.,  2023, \mn@doi [ApJ Lett.]
  {10.3847/2041-8213/acd6a0}, \href
  {https://ui.adsabs.harvard.edu/abs/2023ApJ...950L..10S} {950, L10}

\bibitem[\protect\citeauthoryear{{Taam}}{{Taam}}{1980}]{taam80-1}
{Taam} R.~E.,  1980, \mn@doi [ApJ] {10.1086/158509}, \href
  {https://ui.adsabs.harvard.edu/abs/1980ApJ...242..749T} {242, 749}

\bibitem[\protect\citeauthoryear{{Tanikawa}, {Nomoto}, {Nakasato}  \&
  {Maeda}}{{Tanikawa} et~al.}{2019}]{tanikawaetal19-1}
{Tanikawa} A.,  {Nomoto} K.,  {Nakasato} N.,   {Maeda} K.,  2019, \mn@doi [ApJ]
  {10.3847/1538-4357/ab46b6}, \href
  {https://ui.adsabs.harvard.edu/abs/2019ApJ...885..103T} {885, 103}

\bibitem[\protect\citeauthoryear{{Thielemann}, {Nomoto}  \&
  {Yokoi}}{{Thielemann} et~al.}{1986}]{thielemannetal86-1}
{Thielemann} F.~K.,  {Nomoto} K.,   {Yokoi} K.,  1986, A\&A, \href
  {https://ui.adsabs.harvard.edu/abs/1986A&A...158...17T} {158, 17}

\bibitem[\protect\citeauthoryear{{Vennes}, {Nemeth}, {Kawka}, {Thorstensen},
  {Khalack}, {Ferrario}  \& {Alper}}{{Vennes} et~al.}{2017}]{vennesetal17-1}
{Vennes} S.,  {Nemeth} P.,  {Kawka} A.,  {Thorstensen} J.~R.,  {Khalack} V.,
  {Ferrario} L.,   {Alper} E.~H.,  2017, \mn@doi [Sci]
  {10.1126/science.aam8378}, \href
  {https://ui.adsabs.harvard.edu/abs/2017Sci...357..680V} {357, 680}

\bibitem[\protect\citeauthoryear{{Wang} \& {Han}}{{Wang} \&
  {Han}}{2012}]{wang+han12-1}
{Wang} B.,  {Han} Z.,  2012, \mn@doi [New Astron. Rev.]
  {10.1016/j.newar.2012.04.001}, \href
  {https://ui.adsabs.harvard.edu/abs/2012NewAR..56..122W} {56, 122}

\bibitem[\protect\citeauthoryear{{Werner}, {Reindl}, {Rauch}, {El-Badry}  \&
  {B{\'e}dard}}{{Werner} et~al.}{2024a}]{werneretal24-1}
{Werner} K.,  {Reindl} N.,  {Rauch} T.,  {El-Badry} K.,   {B{\'e}dard} A.,
  2024a, \mn@doi [A\&A] {10.1051/0004-6361/202348286}, \href
  {https://ui.adsabs.harvard.edu/abs/2024A&A...682A..42W} {682, A42}

\bibitem[\protect\citeauthoryear{{Werner}, {El-Badry}, {G{\"a}nsicke}  \&
  {Shen}}{{Werner} et~al.}{2024b}]{werneretal24-2}
{Werner} K.,  {El-Badry} K.,  {G{\"a}nsicke} B.~T.,   {Shen} K.~J.,  2024b,
  \mn@doi [A\&A] {10.1051/0004-6361/202451635}, \href
  {https://ui.adsabs.harvard.edu/abs/2024A&A...689L...6W} {689, L6}

\bibitem[\protect\citeauthoryear{{Wong}, {White}  \& {Bildsten}}{{Wong}
  et~al.}{2024}]{wongetal24-1}
{Wong} T. L.~S.,  {White} C.~J.,   {Bildsten} L.,  2024, \mn@doi [ApJ]
  {10.3847/1538-4357/ad6a11}, \href
  {https://ui.adsabs.harvard.edu/abs/2024ApJ...973...65W} {973, 65}

\makeatother
\end{thebibliography}

%\begin{thebibliography}{99}
%\bibitem[]{Author2012}
%\end{thebibliography}

%%%%%%%%%%%%%%%%%%%%%%%%%%%%%%%%%%%%%%%%%%%%%%%%%%

%%%%%%%%%%%%%%%%% APPENDICES %%%%%%%%%%%%%%%%%%%%%

%%%%%%%%%%%%%%%%%%%%%%%%%%%%%%%%%%%%%%%%%%%%%%%%%%

% Don't change these lines
\bsp	% typesetting comment
\label{lastpage}
\end{document}